\documentclass[10pt,journal,cspaper,compsoc]{IEEEtran}
\usepackage{bbding}

\usepackage{graphicx}
\usepackage{amssymb}
\usepackage{multirow}
\usepackage{caption}
\usepackage{subfigure}
\usepackage{array}
\usepackage[cmex10]{amsmath}
\usepackage[ruled]{algorithm2e}
\usepackage{tabularx}
\usepackage{booktabs}
\usepackage{mdwmath}
\usepackage{mdwtab}
\usepackage{cite}

\begin{document}
\title{Ghost-in-the-Wireless: Energy Depletion Attack on ZigBee}

\author{Devu Manikantan Shila,
        Xianghui Cao,~\IEEEmembership{Member,~IEEE,}
        Yu Cheng,~\IEEEmembership{Senior Member,~IEEE,}
        Zequ Yang,
        Yang Zhou,
        and~Jiming Chen,~\IEEEmembership{Senior Member,~IEEE}% <-this % stops a space
\IEEEcompsocitemizethanks{\IEEEcompsocthanksitem D. Shila is with United Technologies Research Center,
              Hartford, CT 06108 USA.\hfil\break% \protect\\
\IEEEcompsocthanksitem X. Cao and Y. Cheng are with Department of Electrical and Computer Engineering, Illinois Institute of Technology, Chicago, IL 60616, USA. Email: \{xcao10,cheng\}@iit.edu.\hfil\break% \protect\\
\IEEEcompsocthanksitem Z. Yang, Y. Zhou and J. Chen are with Department of Control, Zhejiang University, Hangzhou 310027, China.}
\thanks{}}

\IEEEcompsoctitleabstractindextext{%

\begin{abstract}
ZigBee has been recently drawing a lot of attention as a promising solution for ubiquitous computing. The ZigBee devices are normally resource-limited, making the network susceptible to a variety of security threats. This paper presents a severe attack on ZigBee networks termed as {\em ghost}, which leverages the underlying vulnerabilities of the IEEE 802.15.4 security suites to deplete the energy of the devices. We manifest that the impact of {\em ghost} is severe as it can reduce the lifetime of devices from years to days and facilitate a variety of threats including denial of service and replay attacks. We highlight that merely deploying a standard suite of advanced security techniques does not necessarily guarantee improved security, but instead might be leveraged by adversaries to cause severe disruption in the network. We propose several recommendations on how to localize and withstand the {\em ghost} and other related attacks in ZigBee networks. Extensive simulations are provided to show the impact of the {\em ghost} and the performance of the proposed recommendations. Moreover, physical experiments also have been conducted and the observations confirm the severity of the impact by the {\em ghost} attack. We believe that the presented work will aid the researchers to improve the security of ZigBee further.

\end{abstract}

}

\maketitle

\IEEEdisplaynotcompsoctitleabstractindextext

\IEEEpeerreviewmaketitle

\section{Introduction}
\IEEEPARstart{D}{ue} to its expandability, low cost, ease-of-use, and minimal maintenance, IEEE 802.15.4 based ZigBee,  has been recently drawing a lot of attention to become the most prevalent solution for ubiquitous computing in everyday life. Since the origin of ZigBee, ZigBee alliance has essentially targeted their efforts on building a global wireless language for myriad of everyday devices such as light switches, thermostats, smart devices, remote controls as well as more complex sensor devices found abundantly in the health care, commercial building and industrial automation sectors \cite{poslad2011ubiquitous,alemdar2010wireless,he2011optimal,khan2012ubiquitous,chebrolu2013esense}.

Most of the applications over ZigBee are, however, security sensitive. For instance, it is envisioned that in a smart grid network, automated smart meters will exchange information about the energy consumption of homes to utility companies in a timely fashion. If such information is delivered in a plain-text manner, others could retrieve sensitive private information about the home residents such as their living habits and the time they are not at home. Malicious ones can also inject false energy use information to interrupt the billing system \cite{neuman2011mediating}. The IEEE 802.15.4 standard addresses the security requirements through a medium access control (MAC) layer package, providing fundamental security services ranging from data confidentiality, data integrity to replay protection \cite{IEEE802.15.4}. Despite that the standard provides these basic services, Sastry \textit{et al.} outlined a number of security problems and pitfalls when using the IEEE 802.15.4 specification, especially pertaining to the initialization vector management, key management and integrity protection \cite{Sastry}, and they also suggested several recommendations to improve the security posture of the specification. Zheng \textit{et al.} presented more attacks on the physical and MAC sub-layers, including jamming, capture and tampering, exhaustion, collision and unfairness \cite{zheng2006toward}. Besides research efforts, nowadays, off-the-shelf attack toolkits like KillerBee~\cite{killerbee} are available that can be leveraged even by a novice adversary to explore and exploit the security of ZigBee networks. Using KillerBee and an IEEE 802.15.4 compatible radio interface, an adversary can carry out several attacks ranging from surreptitious eavesdropping to traffic injection with a little or no effort.

Markedly, people need to have a solid understanding of the security performance of ZigBee before positioning it as a major player in the market of ubiquitous computing. In this paper, our in-depth analysis of the ZigBee standard identified a potential flaw related to sending security headers in clear text. These security headers are treated as critical parameters in the specification to provide semantic security and replay protection. A key question is: {\em what might be the consequences if a malicious one masquerades as a trusted device by crafting the security headers?} IEEE 802.15.4 provides protection to integrity related attacks by including a cryptographically secure checksum (aka message integrity code (MIC)) with each message sent to a recipient. When an adversary crafts an invalid security header without knowing the key generating the MIC, although the integrity attack fails, \emph{the recipient device in fact expends certain amount of energy receiving and processing those bogus messages}. In particular, it is shown that the security computing energy can be far from ignorable \cite{xiao2006mac, doomun2007energy}. If an intelligent adversary sends a number of such crafted messages to the victim device, a significant amount of energy will be used leading to battery depletion of the device.

There are studies on jamming attacks, MAC misbehavior, sleep deprivation via power attacks, and link layer exhaustion attacks in wireless networks \cite{zheng2006toward,li2010optimal,tang2014real}. Smart adversaries may attack the protocols in a deeper level, for instance Temporal Key Integrity Protocol (TKIP) MIC attack in IEEE 802.11 networks in which an attacker decodes the payload one byte at a time by using multiple replays and observing the response over the air on MIC failures \cite{lashkari2009survey}, TCP SYN attack in which an attacker sends a chain of SYN requests to a victim system in an attempt to consume enough server resources and launch denial of service (DoS) attack \cite{wang2002detecting} against public key cryptographic operations in sensor networks \cite{dong2013providing}. These attacks thwart the legitimate devices from using the medium and thereby are able to consume large amounts of victim devices' energy, referring to \cite{survey,sleep} and the references therein. As opposed to these efforts, this paper demonstrates a novel and severe attack termed as {\em ghost-in-the-wireless} (aka {\em ghost}) on commercial ZigBee networks (with symmetric keys), in which a malicious one constructs bogus messages to lure the receiver to do superfluous security-related computations to intentionally deplete the energy of devices. The aftermath of the {\em ghost} attack is perilous as it will cut back the lifetime of devices from years to days (to be demonstrated by our simulation and experimental results) and further facilitate an adversary to execute a variety of after-depletion threats like denial of service, replay attack and loss of confidentiality.

In the literature, there also resource depletion attacks in many wireless networks. For example, adversaries can repeatedly send connection requests to exhaust the energy of implantable medical devices \cite{hei2010defending}. In ad hoc sensors networks, an adversary can drain the energy of sensors by purposely sending messages to construct artificial routing paths or introduce loops to the routing process of legitimate sensors \cite{vasserman2013vampire}. However, how to execute such attacks on ZigBee is unknown yet. Moreover, the seriousness of {\em ghost} attack on commercial ZigBee networks, where devices that run on batteries have stringent power constraints, has not been explored.

In this paper, we exploit the ZigBee security headers to launch the {\em ghost} attack. We highlight the fact that merely deploying standard advanced security protocols does not necessarily ensure improved security, but instead might be leveraged by attackers to cause severe disruption in the network. Moreover, we provide qualitative analysis and quantitative simulations as well as experiments to demonstrate the severity of the {\em ghost} attack. We believe that our work in this paper will aid researchers to further improve the security posture of energy-constrained wireless networks.

In summary, this paper has five-fold contributions:
\begin{itemize}
\item We propose a novel and severe attack on ZigBee networks, termed as {\em ghost}, which exploits the underlying vulnerabilities of the IEEE 802.15.4 security suites to cause an intentional energy failure of devices.

\item We theoretically analyze the impact of the {\em ghost} attack on the victim node's lifetime, and develop an analytical model to quantify the impact of DoS attack (induced from {\em ghost}) on the throughput of a multi-hop chain network.

\item We propose a three-phase algorithm to detect and localize the attacker based on network flow variations. We also propose several recommendations on how to withstand the {\em ghost} and its related attacks in ZigBee networks.

\item Extensive computer simulation results are presented to demonstrate the impact of the {\em ghost} attack and the efficiency of the proposed solutions. Particularly, we show that the {\em ghost} attack can not only reduce the device lifetime from years to days by depleting energy, but also lead to the violation of the three basic principles of security, confidentiality, integrity and availability.

\item We validate the effectiveness of {\em ghost} attack by conducting physical experiments on both a single-hop and a multi-hop ZigBee networks, where each node is equipped with CC2420 RF Transceiver that is compliant with the 2.4GHz IEEE 802.15.4 standard.
\end{itemize}

The rest of the paper is organized as follows. Section \ref{background} reviews the IEEE 802.15.4 security architecture. Section \ref{ghost} presents the {\em ghost} attack and some analysis. Section \ref{otherattack} discusses some other severe attacks stemming from the {\em ghost} attack while Section \ref{sec:countermeasures} proposes the corresponding countermeasures. Section \ref{simulation} presents the simulation results to demonstrate the impact of {\em ghost} and {\em ghost}-related attacks and the proposed solutions. Section \ref{experiment} presents the physical experiment results. Finally, Section \ref{conclude} gives the conclusion remarks.

\section{Security Architecture}\label{background}

In this section, we mainly pay our attention to the security services provided by the IEEE 802.15.4 MAC layer \cite{IEEE802.15.4}. The standard particularly provides four basic security services for use by the higher layer applications: access control, data integrity, data confidentiality, and sequential freshness for replay protection, each of which is described briefly in the sequel. The MAC layer is responsible for providing security services on specified incoming and outgoing frames when requested to do so by the higher layers. The higher layer (i.e., the application layer) indicates its choice of the security suite by setting the {\em security control field} in the {\em auxiliary security header} of the message, which identifies 8 candidate security levels ranging from 000 to 111, as shown in Table \ref{tab_level}. The security level configuration can be adjusted message by message. The absence of security parameters indicates ``no security by default''. We urge the interested readers to refer the excellent write up in \cite{Sastry} for details about the security architecture of the standard.

\renewcommand\arraystretch{1.5}
\begin{table}[!ht]
\scriptsize
    \center
    \caption{Security suites in IEEE 802.15.4.}
    %\footnotesize
    \begin{tabular}{cccc}
    \toprule
    \textbf{Security Level/Id} & \textbf{Security Suite} & \textbf{Confidentiality} & \textbf{Integrity}\\
    \hline
    000 & None & \XSolidBrush  & \XSolidBrush\\
    \hline
    001 & AES-CBC-MAC-32 & \XSolidBrush & \Checkmark \\
%    \hline
    010 & AES-CBC-MAC-64 & \XSolidBrush & \Checkmark \\
%    \hline
    011 & AES-CBC-MAC-128 & \XSolidBrush & \Checkmark\\
    \hline
    100 & AES-CTR & \Checkmark & \XSolidBrush\\
    \hline
    101 & AES-CCM-32 & \Checkmark & \Checkmark \\
%    \hline
    110 & AES-CCM-64 & \Checkmark & \Checkmark \\
%    \hline
    111 & AES-CCM-128 & \Checkmark & \Checkmark\\
    \bottomrule
    \end{tabular}\label{tab_level}
\end{table}

{\bf Access Control}:
The IEEE 802.15.4 MAC layer protocol prevents unauthorized devices from participating in the network by maintaining a list of valid devices, commonly known as access control list (ACL). For each incoming message, the receiving device checks the source address against the list of valid addresses in the table. If there is a match, the message is either accepted or forwarded to the next hop, otherwise it is dropped. Although such an access control mechanism can keep out the unauthorized parties from participating in the network, a number of issues emerge such as the spoofing attacks (data integrity issue) where an adversary masquerade as a valid user through crafting messages (e.g., source address) to bypass the ACL checks.

{\bf Data Integrity}:
The standard resolves data integrity issue by including a message integrity code, which is computed by applying a hash function over the message and pre-shared secret key (aka the symmetric key) \cite{zheng2004will}. The MIC tag along with the message is then sent to the receiver. Upon reception, the receiver can validate the integrity by checking whether the received MIC tag can be regenerated using the same hash function, the shared symmetric key, and the received message. If positive, the integrity is considered as maintained, i.e., both the message and the MIC tag were not modified; otherwise, the received message will be discarded. The standard provides data integrity services through AES-CBC-MAC and AES-CCM with three possible lengths of the MIC tag, i.e., 32 bits, 64 bits, or 128 bits.

{\bf Data Confidentiality}:
Underpinning the goal of confidentiality are the encryption schemes. Besides data encryption, the semantic security is needed to ensure that the attacker cannot learn even the partial information about the messages that have been encrypted. A common approach to realizing semantic security is to leverage a unique {\em nonce}, typically a counter or random value, for each invocation of the encryption algorithm. The main purpose of a nonce is to add discrepancy to the encryption process when there is little or no discrepancy in the set of messages. Since the receiver should rely on the same nonce to decrypt the messages, nonces are typically sent in the same message with the encrypted data in plain text, without keeping it secret. {\em This work is concerned with the malicious ones manipulating those nonces sent in plain text.} In the sequel, we will show that how an adversary can leverage the plain-text nonce to launch the {\em ghost} attack. To provide semantic security, both algorithms of AES-CTR and AES-CCM use a 13-bytes nonce, which consists of an 8-bytes source address and a 5-bytes counter (that comprises the frame counter (4 bytes) and the security control field (1 byte) defined in the auxiliary security header).

{\bf Sequential Freshness for Replay Protection}:
Although data confidentiality and data integrity can prevent the network from a variety of known threats such as eavesdropping and spoofing, these schemes cannot protect the network from replay attacks. With the IEEE 802.15.4 specification, the sender usually assigns a monotonically increasing frame counter to each message and the receiver rejects those messages with smaller sequence numbers than it has already seen. The efficiency of this scheme clearly depends on the amount of time it will take the frame counters to roll over. To avoid roll over issues, a 32-bits counter is used in the IEEE 802.15.4. In addition to replay protection, the frame counters are considered as an important input to the construction of nonces for providing semantic security.

\section{Ghost-in-the-wireless}\label{ghost}
In this work, we consider a ZigBee network in which devices are statically deployed and communicate with one another to form a multi-hop wireless backbone. One or more devices serve as the coordinator (or gateways) and provide services for the entire ZigBee network. This work assumes that all the devices are collaborative, behave normally, and follow the algorithms correctly by sending the messages periodically to the coordinator. Notice that there are already works that talk about detecting the nodes that does not obey algorithms \cite{DEVU}. Similar to the standard, the cryptographic mechanism presented in this work is based on the symmetric-key cryptography and uses keys that are provided by higher layer processes. The establishment and maintenance of the keys are outside the scope of this work. As stated in the standard, we assume a secure implementation of the cryptographic operations, and secure and authentic storage of the keying material.

In such an environment, we assume the presence of single or multiple {\em ghost} attackers, equipped with compatible IEEE 802.15.4 radios. As opposed to legitimate devices, the attacker devices have no power or memory constraints and are assumed to span the attacking range over a large number of devices. We assume a three-phase attack model: (a) {\em pre-attack phase} --- in this phase, the attacker learns about the network by surreptitiously eavesdropping, capturing and reverse engineering the messages; (b) {\em Attack phase} --- the attacker leverages the learned information to execute the {\em ghost} attack; and (c) {\em post-attack/depletion phase} --- once the energy of devices are depleted, attacker in this phase executes several other attacks such as replay attack or confidentiality attack, to be discussed in Section~\ref{otherattack}.

\subsection{AES Suites}
ZigBee uses AES \cite{daemen2002design} based security suites to provide fundamental security services like confidentiality, integrity and replay protection. In this section, we will initially shed light on the working of each of these algorithms and then present the proposed {\em ghost} attack.

{\bf AES-CTR for Encryption}.
Messages in this mode are encrypted and decrypted by XORing with the key stream produced by the AES encrypting sequential counter block values. Let $O={O_1,O_2, \cdots, O_n}$  denote the output keystream block. To encrypt a payload with AES-CTR, the encrypter partitions the plaintext, $P$, into $n$ 128-bit blocks ${P_1, P_2 \cdots, P_n}$; notice that if the last block is not a multiple of 128 bits, zeros are padded to it. Each $P$ block is then XORed with a block of the key stream $O$ to generate the corresponding ciphertext, $C={C_1, C_2, \cdots, C_n}$. To avoid reuse of the same output stream $O$, AES-CTR requires the sender (encrypter) to generate a unique key stream per-block per message. The decryption process is similar to encryption process.

{\bf AES-CBC-MAC for Authentication}.
A message is authenticated by splitting the input $I$ into $n$ 128 bit blocks, with necessary padding. Let $I={I_1, \cdots, I_n}$ and $O={O_1, \cdots, O_n}$ denote the input and its corresponding output block, where $O=CIPH(k)[I]$ and $CIPH(k)[I]$ is the invocation of AES algorithm on the  input block using the secret key $k$. The CBC-MAC mode is then defined as: $O_1=CIPH(k)[I_1]$, $O_2=CIPH(k)[I_2 \oplus O_1]$, $\cdots$, $O_n=CIPH(k)[I_n \oplus O_{n-1}]$ and the final block is the MIC.

{\bf AES-CCM for Encryption and Authentication}.
This mode consists of two steps: computing the MIC tag using CBC-MAC and encrypting the message concatenated with the MIC tag using CTR mode. Let $P=P_1, P_2, \cdots, P_n$ and $O=O_0, O_1, O_2, \cdots, O_n$ denote the plaintext and the output keystream blocks. Then the CCM mode is defined as follows:  $P_1 \oplus O_1, P_2 \oplus O_2, \cdots P_n \oplus O_n || MIC \oplus O_0$.

\subsection{Attack Phase}
AES-CTR, AES-CBC-MAC and AES-CCM depend on the encryptor to generate a unique  keystream per message ($O$) to provide semantic security. This task is accomplished by leveraging a 16-bytes unique counter constructed from the fields in the message intended for the destination. The 16-bytes counter consists of a 2-bytes static flags field, a 13-bytes nonce field (comprising the sender address, the frame counter and the security level field), and a 1-byte block counter that numbers the 16-bytes blocks within the message. In addition to semantic security, the frame counters are used by the security suite to enable replay protection as well. When receiving a message, the recipient compares the frame counter seen in the incoming packet to the highest value stored in the ACL table. If the incoming packet has a larger counter value than the stored one, the message is accepted and the new counter is used to update the ACL table; otherwise, the message is rejected. With a 4-bytes frame counter, an adversary can carry out a replay attack only after $2^{32}$ frames, which is considered cryptographically secure in practice.

In the IEEE 802.15.4 protocol, those message fields used to construct the 16-bytes counter need to be communicated in plain text by the sender. Suppose that a malicious one injected messages into the network with increasing frame counters. According to the standard, the receiving device will accept it as the incoming message has a larger counter value than the last observed one. However, on decrypting the message, receiver will understand that the message is corrupted by computing the checksum or the MIC and finally, will drop it. Although the integrity attack fails, the recipient device in fact expends a valuable amount of energy accepting and processing those bogus messages.  If an intelligent attacker sends a number of such crafted bogus messages to the victim device to lure the receiver to do the superfluous security-related computations, a significant amount of energy will be spent leading to battery depletion. We therefore term this attack as {\em ghost-in-the-wireless}. Although there are resource depletion attacks, we show by {\em ghost} how such an attack can be executed in ZigBee. Our simulation results show that through {\em ghost} attack, one can cut back the lifetime of devices from years to days.

\subsection{Analysis}\label{sec:analysis}
We present an analytical model to quantify the effect of a {\em ghost} attacker on a victim device (denoted as $D_i$). %\cite{UIUC}.
Consider that $D_i$ works in a duty-cycling mode, which is a common working mode for low-cost networks such as wireless sensor networks, with duty-cycle equal to $\lambda=\frac{\tau}{T}$ where $\tau$ is the duration of an active period and $T$ is the length of a cycle. Once entering an inactive period (aka sleep period), the device turns off its radio and changes CPU state (i.e., forces CPU into some low-power mode) if no incoming message is being received or decrypted; otherwise, it turns off its radio once the incoming message is finished receiving, and changes CPU state once the message is finished processing. The attacker sends bogus messages with crafted security headers to $D_i$ at a high frequency (say $\rho$ packets per unit time), while only those transmitted during $D_i$'s active period will be effective in depleting its energy. Assume that the attacker messages are of the same size.

For the victim device, denote $T_{rx}$ as the time spent by its radio to receive a bogus message, which depends on message length and data rate. Denote $P_{rx}, P^a_{cpu}, P^i_{cpu}$ and $P^o_{cpu}$ as the power required by its radio when in receiving mode and by its CPU when in active, idle, and low-power modes, respectively. The energy consumption of a wireless device depends on the amount of energy expended on the following states: transmitting, receiving, computing, idle and sleeping states. Based on these states, the energy consumption of $D_i$ is divided  into three costs: communication cost (i.e., when it is receiving) $E_{comm}$, computation cost (i.e., when processing the data) $E_{comp}$, and passive cost (i.e., when it is not involved in communication, say, sleeping) $E_{passive}$.

The computation cost of the device depends on the amount of energy expended for cryptographic operations (including decryption and MIC verification). To compute it, initially, the cost of performing a decryption/verification in unit of ampere-cycle is computed by taking the product of the total number of clock cycles taken by the AES algorithm and the average current drawn by each CPU cycle. The total energy cost is then computed by dividing the  ampere-cycles by the clock frequency in cycles/second of a processor and further multiplying the result with the processor's operating voltage. Thus the computation cost for the decryption and verification of one message is $T_{dec}P^a_{cpu}$, where $T_{dec}$ denotes the time required for decryption and verification of a bogus message, which depends on the specific security suite used. Taking the CPU cost during the packet receiving period into account, we have
\begin{align}
E_{comp} =& n_p\left(T_{dec}P^a_{cpu}+ T_{rx}P^i_{cpu}\right)
\end{align}
where $n_p$ is the number of messages that will be processed during an active period. $n_p\approx\left\lceil\frac{\tau}{\max\{T_a,\frac{1}{\rho}\}}\right\rceil$, where $T_a \triangleq T_{dec} + T_{rx}$ is the total time to receive and process one bogus message.

As we are concerned about the amount of energy spent by the victim device $D_i$ in receiving the bogus packets from the {\em ghost} attacker, the communication cost will be mainly determined by the receiving cost and is
\begin{align}
\nonumber E_{comm} \approx &  \left\{\begin{array}{ll}
        n_p T_{a} P_{rx}, \quad & \textrm{if } n_p T_{a}\geq \tau \\
        \tau P_{rx}, \quad & \textrm{otherwise}
    \end{array}\right.\\
         =& \max\{n_p T_{a},\tau\} P_{rx}
\end{align}

The passive cost includes sleep cost and idle cost between the completion time for processing a current message and the arriving time of the next message. Hence,
\begin{equation}
   E_{passive} = \left\{\begin{array}{l}
        (\tau-n_p T_a)P^i_{cpu} + (T-\tau)P^o_{cpu}, \\
            \qquad\qquad\qquad\qquad \text{ if } n_p T_a<\tau\\
        (T-n_p T_a)P^o_{cpu}, \quad\text{ otherwise}.
        \end{array}\right.
\end{equation}

Then, the energy consumption of the victim device receiving and processing a message from the {\em ghost} attacker is computed as $E_{p} = E_{comm} + E_{comp} + E_{passive}$.

\subsubsection{Number of Messages Leading to Depletion}
Let $E_{residual}$ be the amount of energy available to the device and $E_{threshold}$ be the threshold level below which the device fails to participate in the network, then we have the following condition to be true for the {\em ghost} attack to be successful:
\begin{equation}
E_{residual}-mE_{p} \leq E_{threshold}
\label{eq_ghost}
\end{equation}
where $m$ is the number of bogus messages for victim device. From eqn.(\ref{eq_ghost}), it follows that the {\em ghost} attacker should send at least $m \geq \frac{E_{residue}-E_{threshold}}{E_p}$ packets to successfully deplete the energy.

\subsubsection{Lifetime Reduction}
Let $E_0$ be the initial energy of the victim device. If no attacker presents, its lifetime is $L_0=\frac{E_0}{E_{p,0}}$ where
\begin{equation}
    E_{p,0} = \tau (P_{rx}+P_{cpu}^i),
\end{equation}
if we neglect the sleep cost. If $n_p T_a\geq \tau$, based on the above analysis, we have $E_p = n_p(T_{dec}P^a_{cpu}+ T_{rx}P^i_{cpu}) + n_p T_{a} P_{rx}$. Thus, the ratio of $L_0$ over the victim device's lifetime under {\em ghost} attack becomes
      \begin{align}
      \nonumber       \frac{L_0}{L} =& n_p T_{a} \frac{\frac{T_{dec}}{T_a}P^a_{cpu}+ \frac{T_{rx}}{T_a}P^i_{cpu} + P_{rx}}{\tau (P_{rx}+P_{cpu}^i)}\\
                      =& \frac{n_p T_{a}}{\tau} \left(1+\frac{T_{dec}}{T_a} \frac{P^a_{cpu}-P^i_{cpu}}{P_{rx}+P_{cpu}^i}\right) > \frac{n_p T_{a}}{\tau}
      \end{align}
That is $\frac{L}{L_0} < \frac{\tau}{n_p T_{a}}$. Similarly, if $n_p T_a< \tau$, %we can obtain that
      \begin{align}
             \frac{L_0}{L} =& 1+\frac{n_pT_{dec}}{\tau} \frac{P^a_{cpu}-P^i_{cpu}}{P_{rx}+P_{cpu}^i}
      \end{align}

Above two equations clearly show that the lifetime of the victim device is reduced to some extend, and the amount of reduction can be significantly large if $n_p T_a$ is much larger than $\tau$.

\section {Other Attacks Due to Ghost}\label{otherattack}
In this section, we will demonstrate a number of severe attacks stemming from {\em ghost} attack and propose countermeasures to mitigate their impact.

\subsection{Denial of Service (DoS)}\label{sec:dos}
DoS can be easily executed by a {\em ghost} attacker in three ways. (1) {\em DoS due to high computational load on the device}. In a {\em ghost} attack, the adversary sends a number of bogus messages  to quickly deplete the energy of the victim device and thereby, suspend the availability of the services. It turns out that if the network has some  traffic abnormality detection schemes in place, sending such numerous messages in short period of time can be easily caught. To escape from the detection, for instance, {\em ghost} attacker(s) can send messages either at different times or at different addresses to a subset of victim devices in its range. (2) {\em DoS due to MAC misbehavior}: The IEEE 802.15.4 utilizes distributed contention resolution mechanisms such as CSMA/CA for sharing the wireless channel. Under CSMA/CA protocol, only one transmission can happen at any given point in time in a given area and therefore, to achieve this CSMA/CA requires devices to sense the channel for idleness before it can transmit. In such an environment,  if a {\em ghost} attacker continuously sends the traffic to the victim host, all devices within the interference region will be deprived of channel access and services. Moreover, each device has to spend a significant amount of time sensing and waiting to get access to channel, which again will lead to energy depletion. (3) {\em DoS with a post-depletion replay attack}: Such an attack is to be discussed in Section \ref{sec:replayattack:reuse}.

\subsubsection{Analysis}\label{sec:dos:analysis}
\begin{figure}[htbp]
    \centering
    \includegraphics[width=3in, bb=61 278 479 440]{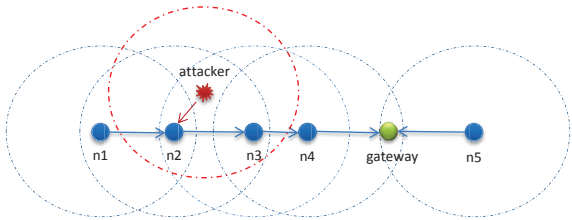}
    \caption{A multi-hop network formed by 5 ZigBee nodes and a gateway, where node 2 and 3 are in the interference range of a {\em ghost} attacker.}
    \label{fig_multihop}
\vspace{-3mm}
\end{figure}

To illustrate the effect of DoS attack on the network throughput performance, we conduct mathematical analysis based on a simple multi-hop network as shown in Fig. \ref{fig_multihop}. All the nodes apply the CSMA/CA protocol for channel access. Assume each node randomly generates packets (of the same length $L$) at the same rate $\lambda$ packets/slot, where a slot is the backoff slot in IEEE 802.15.4 standard. For ease of analysis, we use the following CSMA/CA parameters: the minimum and maximum backoff exponents are the same as $macBE$, CW=1, i.e., a node will start to transmit packet if it finds the channel is idle in previous slot. The nodes send packets to the gateway through the paths shown in this figure. A node will switch to receiving mode only when its MAC buffer is empty. The gateway node keeps in receiving mode and listening packets from others. The {\em ghost} attacker sends bogus packets at a constant rate $p_{att}$ independent of the channel state.

For each node, define $\tau_i$, $\alpha_i$ and $\rho_i$ as the probabilities that this node conducts a clear channel assessment (CCA), the channel is idle when it conducts a CCA, and that there is at least one packet in its MAC buffer, respectively in a slot. Conditioned on that its buffer is nonempty, a node will transmit a packet with probability $p_i$ and the packet will be successfully received (without collision) by its next-hop node with probability $p^s_i$. We extend the model in \cite{ling2008renewal} for single-hop networks to the multi-hop network shown in Fig. \ref{fig_multihop}. Since the modeling approach is the same for each node, below we shall focus on node 2. Since the backoff exponent is constant, according to \cite{ling2008renewal},
\begin{align}
    \tau_2 = \frac{1}{\bar{b}+1+L\alpha_2},
\end{align}
where $\bar{b}\triangleq\frac{1}{2}(2^{macBE}-1)$ is the average backoff length. When this node conducts a CCA, the channel is idle only when none of its neighbors (including the attacker) is transmitting (i.e., none of them starts transmitting in either of the previous $L$ slots). Therefore,
\begin{align}
%    \alpha_2 = \left[(1-p_{att})(1-\rho_1p_1)(1-\rho_3p_3)\right]^L.
    \alpha_2 = 1-L\left[1-(1-p_{att})(1-\rho_1p_1)(1-\rho_3p_3)\right].
\end{align}
By definition,
\begin{align}
    p_2=\;&\tau_2\alpha_2, \quad \rho_2 =\; \min\left\{\frac{\lambda+\rho_1p_1p^s_1}{p_2},1\right\}.
\end{align}
where $\rho_1p_1p^s_1$ is the rate of packets being received from node 1. When node 2 transmits a packet, the packet will be successfully received by node 3 only if: 1) neither node 1 nor the attacker starts to transmit packet simultaneously as node 2, 2) node 4 does not transmit packet in any slot during the transmission by node 2, and 3) node 3 is in receiving mode. Therefore,
\begin{align}
%    p^s_2 = (1-\rho_1p_1)(1-p_{att})(1-\rho_4p_4)^L(1-\rho_3).
    p^s_2 = (1-\rho_1p_1)(1-p_{att})(1-L\rho_4p_4)(1-\rho_3).
\end{align}

Based on the same approach, the performance of the other nodes can be modeled. Then, the throughput of each node can be calculated.
\begin{align}
\left\{\begin{array}{lcl}
    S_1 &=& \rho_1p_1p^s_1p^s_2p^s_3p^s_4,\\
    S_2 &=& \frac{\lambda}{\lambda+\rho_1p_1p^s_1}\rho_2p_2p^s_2p^s_3p^s_4,\\
    S_3 &=& \frac{\lambda}{\lambda+\rho_2p_2p^s_2}\rho_3p_3p^s_3p^s_4,\\
    S_4 &=& \frac{\lambda}{\lambda+\rho_3p_3p^s_3}\rho_4p_4p^s_4,\\
    S_5 &=& \rho_5p_5p^s_5
\end{array}\right.
\end{align}

\begin{figure}[htbp]
 %\vspace{-1mm}
	\centering
	\subfigure[Throughput---case 1.]{
		\includegraphics[width=1.5in, bb=6 150 559 615]{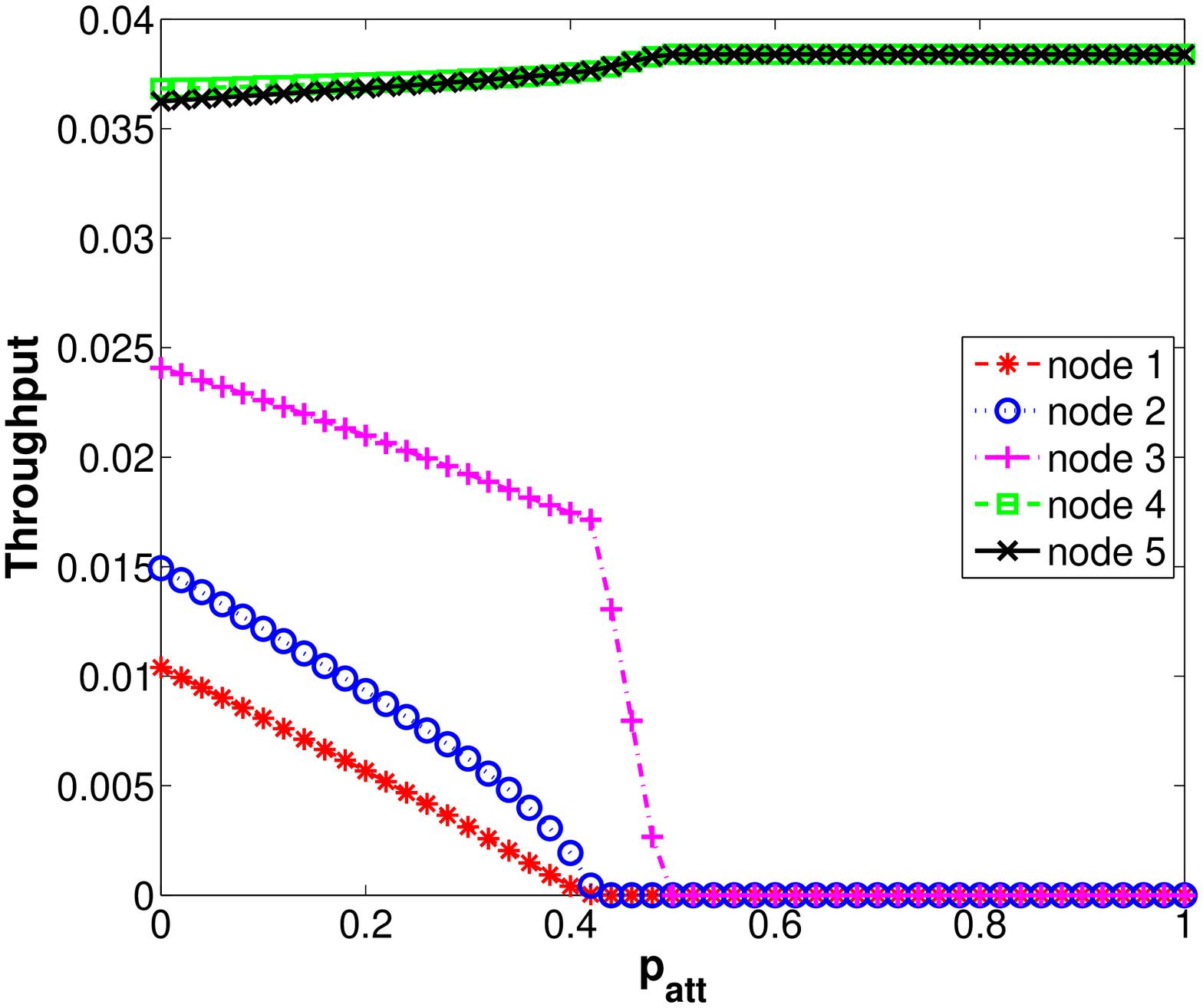}
		\label{fig_multihopS1}
	}
%	\vspace{0.1mm}
	\subfigure[Throughput variation---case 1.]{
		\includegraphics[width=1.5in, bb=4 148 559 615]{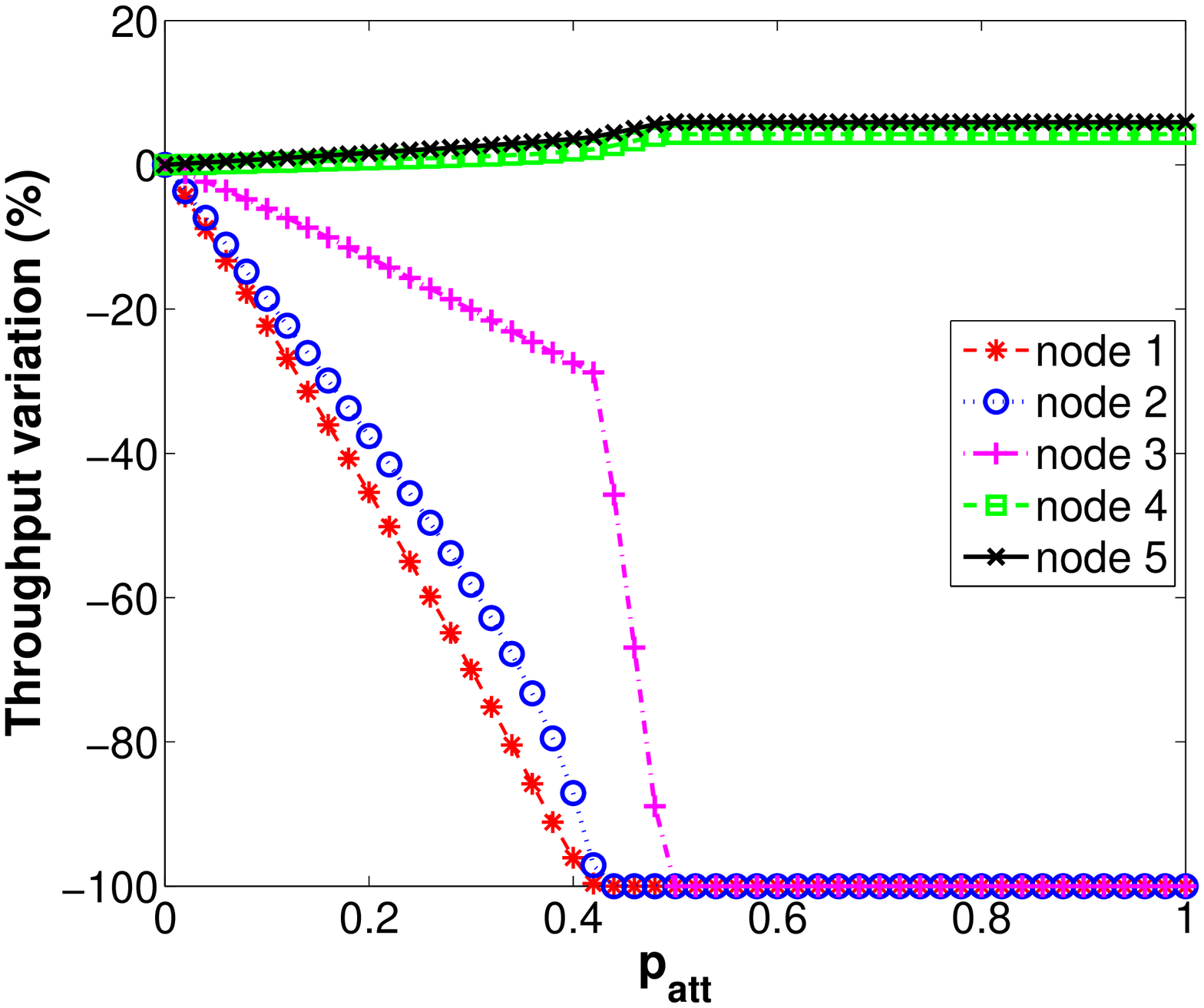}
		\label{fig_multihopSV1}
	}
%    \vspace{0.1mm}
	\subfigure[Throughput---case 2.]{
		\includegraphics[width=1.5in, bb=6 150 559 615]{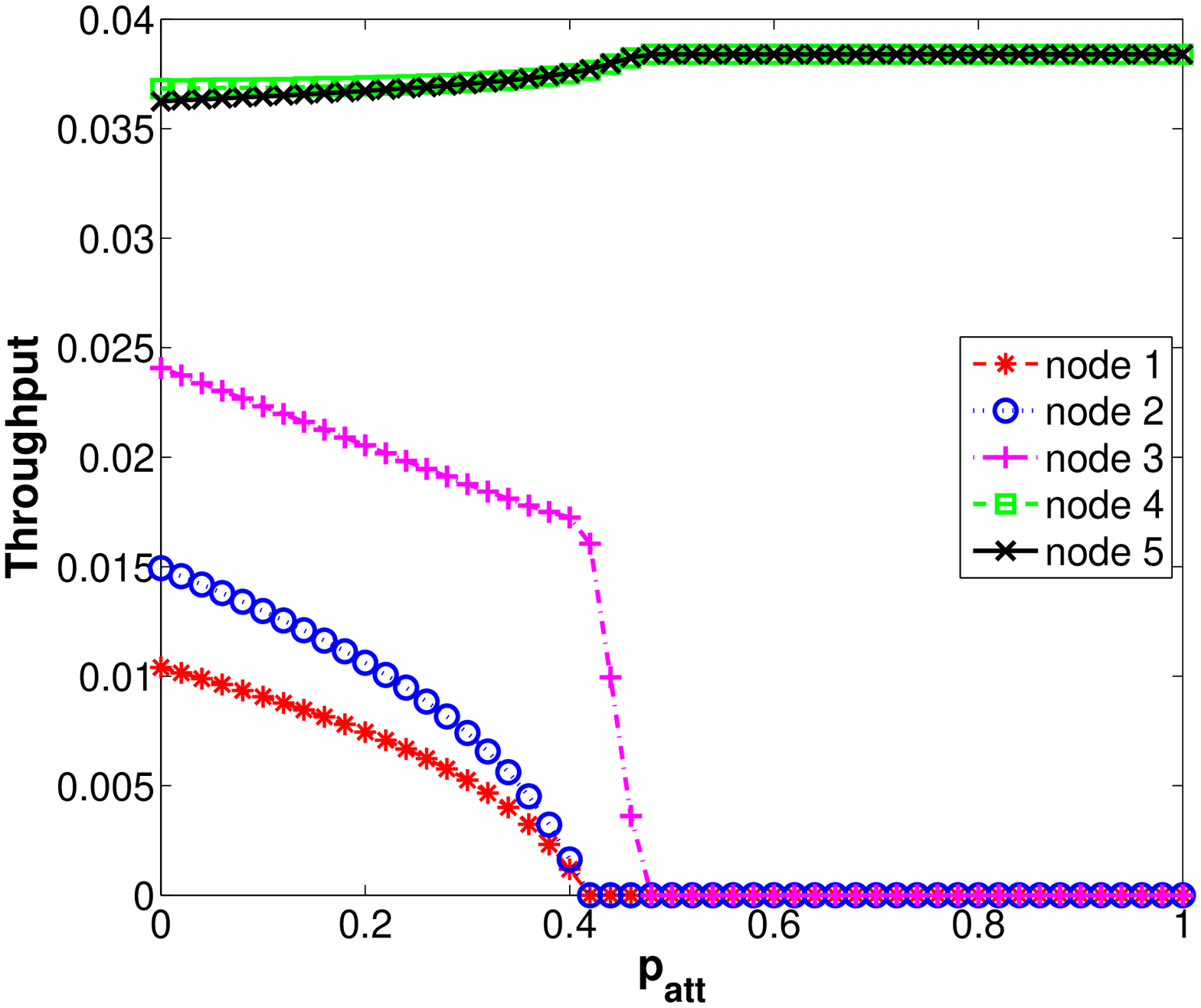}
		\label{fig_multihopS2}
	}
%	\vspace{0.1mm}
	\subfigure[Throughput variation---case 2.]{
		\includegraphics[width=1.5in, bb=4 148 559 615]{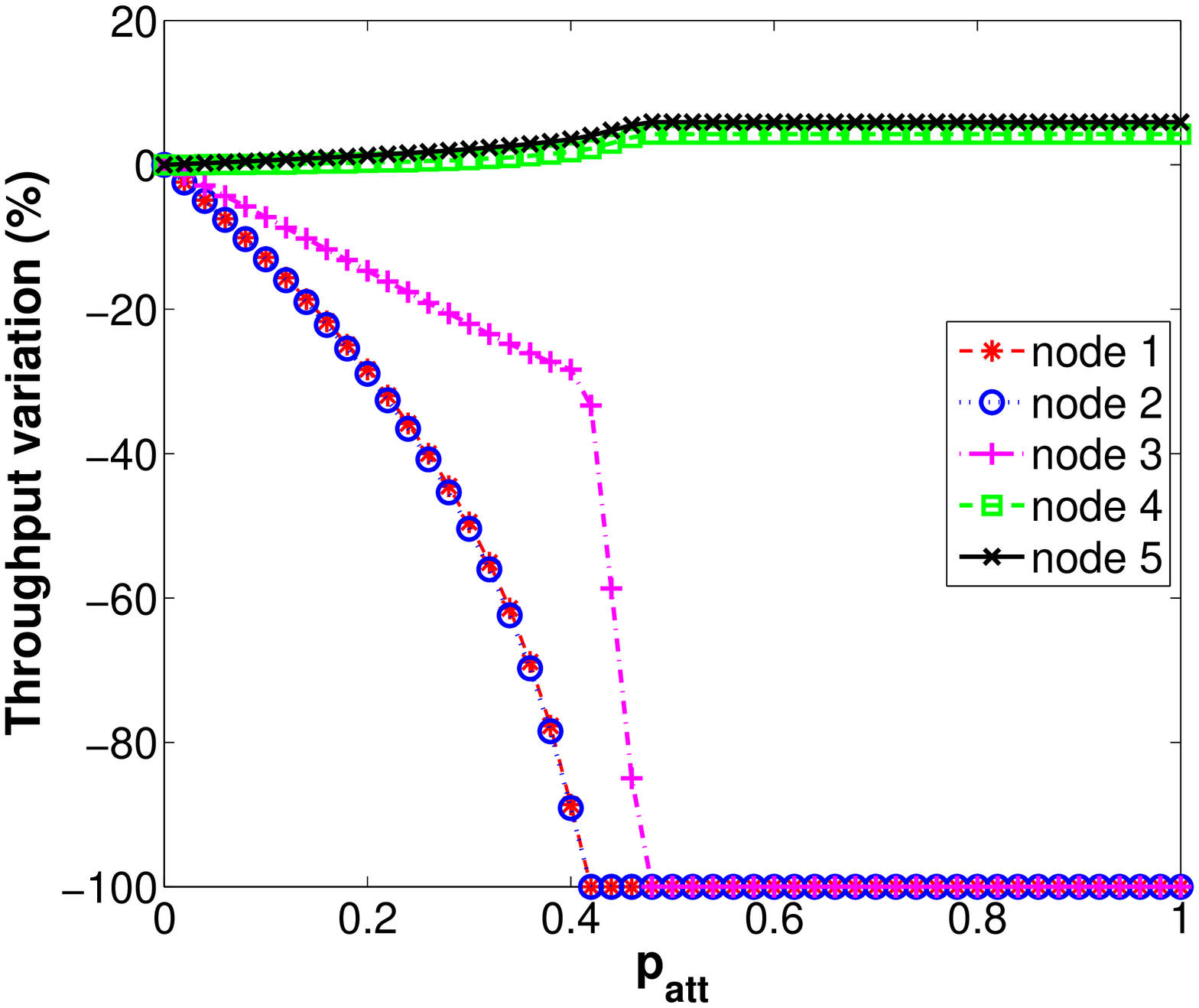}
		\label{fig_multihopSV2}
	}
 %\vspace{-4mm}
	\caption{Network performance under various attack rate. $p_{att}=0$ means the scenario without {\em ghost}.}	
	\label{Fig_multihop}
    \vspace{-3mm}
\end{figure}

We name the scenario shown in Fig. \ref{fig_multihop} as case 1. For comparison, another case (case 2) is also considered in which the attacker's interference range only covers node 3. Numerical results are shown in Fig. \ref{fig_multihop}, where $L=3, macBE=3,\lambda=0.02$. In both cases, due to increase of attack intensity, nodes that are within the interference range of the {\em ghost} have fewer opportunities to send out packets, and their throughput drops quickly. Since packets from node 1 transverse the interfered area, its throughput drops even quicker than that of the nodes within the area. On the other hand, due to less interference from node 3, the throughput of node 4 and 5 whose packet routes do not cross the interfered area may gain some improvement. From Fig. \ref{fig_multihopSV2}, we observe that, even node 2 is a relay of node 1, they experience similar throughput variation when node 4, a common relay of them, is under attack. We take advantage of such observation to develop an attacker localization method in Section \ref{sec:localizaiton}.

\subsection{Post-Depletion Attacks}
What happens when the device encounters an energy depletion? If no specific controls are taken, the device will emerge with a cleared ACL table on reboot and consequently, all the nonces or the frame counters will be reset to the initial value 0. Upon such resetting, we see the plausibility of two attacks:

\subsubsection{Replay Attack due to Frame Counter Reuse}\label{sec:replayattack:reuse}
Replay attack can cause severe problems. For example, medical devices such as fitbits, pacemakers and insulin pumps which replay old messages can lead to erroneous outputs and dangerous events. Below we show that a {\em ghost} can launch replay attack. Consider a {\em ghost} attacker who captures  a certain number of legitimate packets in the beginning (pre-attack phase) and starts depleting the energy of the devices through bogus messages. When the counters are reset to 0 upon restart, the attacker can start replaying those messages. The replay attack can lead to serious results in two aspects. (1) The attack can have the receiver consume obsolete messages, leading to unexpected operations. (2) The replay attack can result in the denial of service. Specifically, if the attacker replays messages with frame counters larger than the current legitimate value maintained in the ACL table, the ACL table will then be updated by such replayed messages. Consequently, in future, when legitimate devices send messages with the actual frame counter, the message will be rejected as it carries a counter value less than what is currently stored in the ACL of the victim device (which was updated by the replayed messages).

We would like to emphasize that the {\em ghost} attack significantly facilitates the chance of launching a replay attack. From the IEEE 802.15.4 specification, it follows that by leveraging a 4-bytes counter, one can execute a replay attack only after $2^{32}$ frames; nonetheless, in this work, we manifest that by executing the {\em ghost} attack, an adversary can use a number of messages which is far less than $2^{32}$ to deplete the energy of a victim device. Upon the restart of this victim device, the attacker can launch the relay attack.

\subsubsection{Loss of Confidentiality due to Nonce Reuse}

According to the standard, messages destined for the next hop devices are encrypted by XORing the plaintext message $P$ with the unique keystream $O$ constructed from the unique nonces, static flags and the pre-shared key. Take the AES-CTR scheme for instance. To encrypt a payload, the encrypter partitions the plaintext, $P$, into $n$ 128-bit blocks ${P_1, P_2 \cdots, P_n}$; if the last block is not a multiple of 128 bits, zeros are padded to it. Each $P$ block is then XORed with a block of the key stream $O$ to generate the corresponding ciphertext, $C={C_1, C_2, \cdots, C_n}=P\oplus O$. If the nonces are reused, then the keystream will repeat for the subsequent messages. Thus, for a different plain text $P'$, the output of the encryptor will be $C'=P' \oplus O$. Assume that the attacker has captured the ciphertexts $C$ and $C'$, then a simple XORing of the two ciphertexts will lead to $C' \oplus C = (P' \oplus O) +(P \oplus O) = P' \oplus P$; such a simple XOR of plain texts can trivially be broken using statistical analysis and the subsequent plaintext messages can be deciphered without the knowledge of pre-shared key,  if the content of any one message can be guessed.

\section{Countermeasures}\label{sec:countermeasures}
In this section, we first consider the problem of {\em ghost} attacker localization. Next, we propose add-on mechanisms to current IEEE 802.15.4 standard to enhance network security against {\em ghost}.

\subsection{Attacker Localization}\label{sec:localizaiton}
Although the specific victim node is able to know the existence of a {\em ghost} attacker by checking whether the number of received bogus messages exceeds a tolerable level, it, under intense attack, may not be able to timely inform others since the channel could be unavailable (occupied by the {\em ghost}) and its CPU may be also busy in processing the bogus messages. Whereas, owing to the induced DoS impact on the network throughput (as demonstrated in Fig. \ref{Fig_dos} in our simulations), other nodes can still be able to detect and localize the attacker by analyzing the network flow changes. Consider the typical cluster based topologies in ZigBee networks. In a cluster, we let the cluster head (CH), the confluence of the flows from the other nodes, to carry out the analysis. As a first attempt to localize {\em ghost} attacker in such a ZigBee network, a three-phase method is proposed as follows. It is difficult, if not impossible, for the CH to determine the number of {\em ghost}s if they collocate or are very close by analyzing their DoS impact. However, knowing the existence of at least one {\em ghost} at a location already satisfies our purpose.

{\bf Phase 1}: Identify suspected victim nodes. The CH applies a moving analysis window to record the throughput of each node\footnote{For dense deployment, to save computation cost in analysis, the CH can select a few paths with the nodes scattering over the area as uniform as possible for analysis.} in the corresponding cluster, and calculates the percentage of throughput variation (denoted as $\Delta S_i$ for node $i$) in real time. Suspected nodes are identified by checking the variation along each path. The observation mentioned in Section \ref{sec:analysis} is utilized to exclude those nodes outside of the attacker's interference area but their packet routes traverse that area. This exclusion operation is controlled by a threshold $\delta'$ as shown in Algorithm \ref{alg}. We denote each path as $P_l=(l_1,l_2,\ldots,l_m)$ with $l_1$ as the source node and $l_m$ (i.e., the head) as the destination node. $\delta$ is used to eliminate throughput variance in normal condition.
\begin{algorithm}[ht]
    $\mathcal{A} \leftarrow \emptyset$: set of suspected victim nodes\;
    \For{all paths $P$}{
    \If{$\Delta S_{l_1}>-\delta$}{
        continue\;
    }
    \For{$i\leftarrow 1,\ldots,m-1$}{
        \If{$\Delta S_{l_i}<-\delta$}{
            \If{$i+1=m$ or $|\Delta S_{l_i}-\Delta S_{l_{i+1}}|>\delta'$}{
                add $l_i$ into $\mathcal{A}$\;
                \If{$|\Delta S_{l_i}-\Delta S_{l_{i-1}}|<\delta'$}{
                    add $l_{i-1}$ into $\mathcal{A}$\;
                }
            }
        }
    }
    }
\caption{Identify suspected victim nodes.}\label{alg}
\end{algorithm}

{\bf Phase 2}: Group suspected victim nodes. We consider a general case that neither the number nor interference ranges of the {\em ghosts} is known to the CH. Two suspected nodes can be considered under the interference of the same attacker and included in the same group if there is a circle that covers both of them but does not cover any other non-suspected nodes. Otherwise, they are considered to be attacked by different attackers and included in different groups. Since an attacker may want to span its attacking range over a large number of nodes, small clusters will not be considered for attacker localization.

{\bf Phase 3}: Calculate the {\em Ghost} location. For each group, suppose there are $s$ suspected nodes and the throughput of each node decreases by $\Delta S_i$ percent. Then, the attacker's location is estimated by the following weighted sum method.
\begin{equation}
    L_{\textrm{\em ghost}} = \sum^{s}_{i=1}\frac{\Delta S_i}{\sum^s_{k=1}\Delta S_k}L_i,
\end{equation}
where $L_i$ is the location of each suspected victim node.

\subsection{Add-on Mechanisms}\label{sec:addonmechanisms}
We propose to add techniques such as {\em blacklisting} to current security services provided by standard, in which each device will maintain a list of devices that are misbehaving. In the case of {\em ghost} attack, if the victim device observes a certain number of messages with bogus security headers, it will add the device to the blacklist and inform the network or the operator about the attack. One plausible issue with this approach is the {\em badmouthing} attack~\cite{DEVU} in which an attacker sends bogus messages from different addresses and causes the victim device to blacklist all its surrounding devices, leading to a temporary disruption or denial of service.

Another approach is to add a second layer of {\em challenge-response scheme} in which the device, after observing a certain number of bogus messages from a specific address or when the energy is restored and the communications are re-initiated, will challenge the attacker with a random number. The solution requires the attacker to include the response to the challenge in the next message for the device. If the attacker is able to respond to  it correctly, it will continue its operation, otherwise, it will inform the network or the operator about the attack. Notice that to include correct response in the message, the attacker needs to know the secret key which is available only to the legitimate parties and is securely stored in the device. On restoration of energy, we also require, as part of the challenge-response scheme,  to establish new keys so that they don't reuse the same nonce twice with the same key. Another solution is to add a timestamp to the protocol and after a reboot, each device is required to update its timestamp by communicating with the controller. There are also works that suggest storing the counter values in non-volatile or flash memory so that even if the energy is lost, the state of the device can be restored. Nevertheless, storing and retrieving values from flash memory is slow and energy inefficient, specifically for energy constrained devices \cite{Sastry}.

\section{Simulations}\label{simulation}
In this section, we present extensive simulation results to demonstrate the impact of the proposed {\em ghost} attack. We develop our simulation codes within the NS-3 environment \cite{NS3}. We consider a ZigBee network with a single coordinator serving as the gateway, $n$ legitimate nodes and a {\em ghost} attacker which injects messages with bogus security headers to drain victim nodes' energy. Each ZigBee node is equipped with an 8 MHz processor. All the nodes apply the non-beacon mode of the IEEE 802.15.4 CSMA/CA protocol for channel access. We adopt the energy consumption model presented in \cite{energymodel}. Specifically, the current drain by each node for its CPU being in active, idle and power-save states are 8.0 $mA$, 3.2 $mA$ and 110 $\mu A$, respectively; the current drain by each node's transceiver for receiving and transmitting are 7.0 $mA$ and 8.5 $mA$, respectively; and the transmission power of each node is 0 dB. We also adopt an accurate Li-Ion battery model (which is provided in NS-3 based on the models proposed in \cite{batterymodel1,batterymodel2}) for each node with the initial capacity set as 2.45 Ah to simulate the energy drain activities. We run the simulation program on a 1.85 GHz PC. Based on the number of instructions, the CPU computation time of the PC is mapped to that of the 8 MHz processors to approximate the computational energy consumption involved in the ZigBee security protocol.

\begin{figure*}[htbp]
	\centering
	\subfigure[]{
		\includegraphics[width=1.6in]{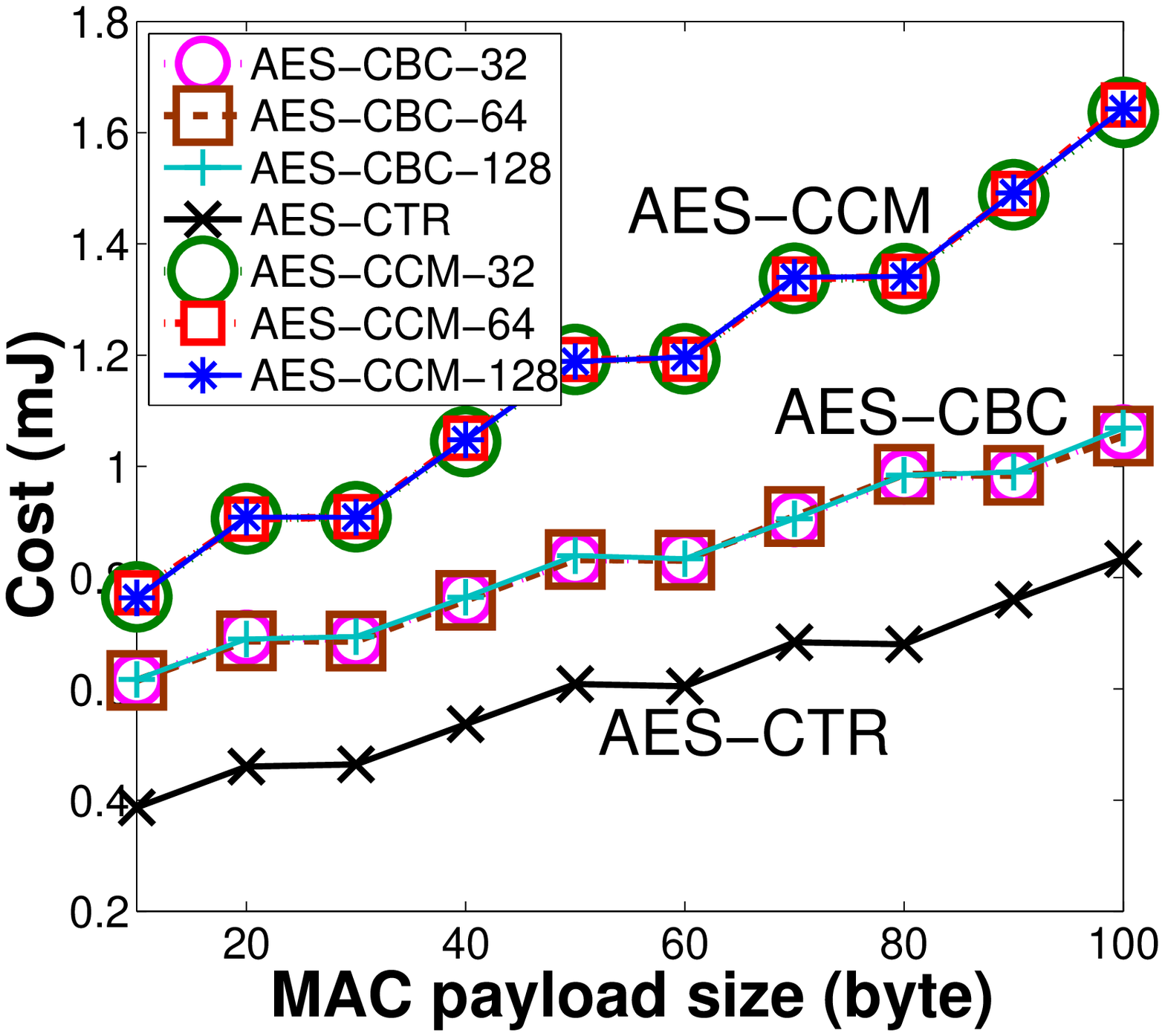}
		\label{fig_pEncCost}
	}
	\hspace{-3mm}
	\subfigure[]{
		\includegraphics[width=1.6in]{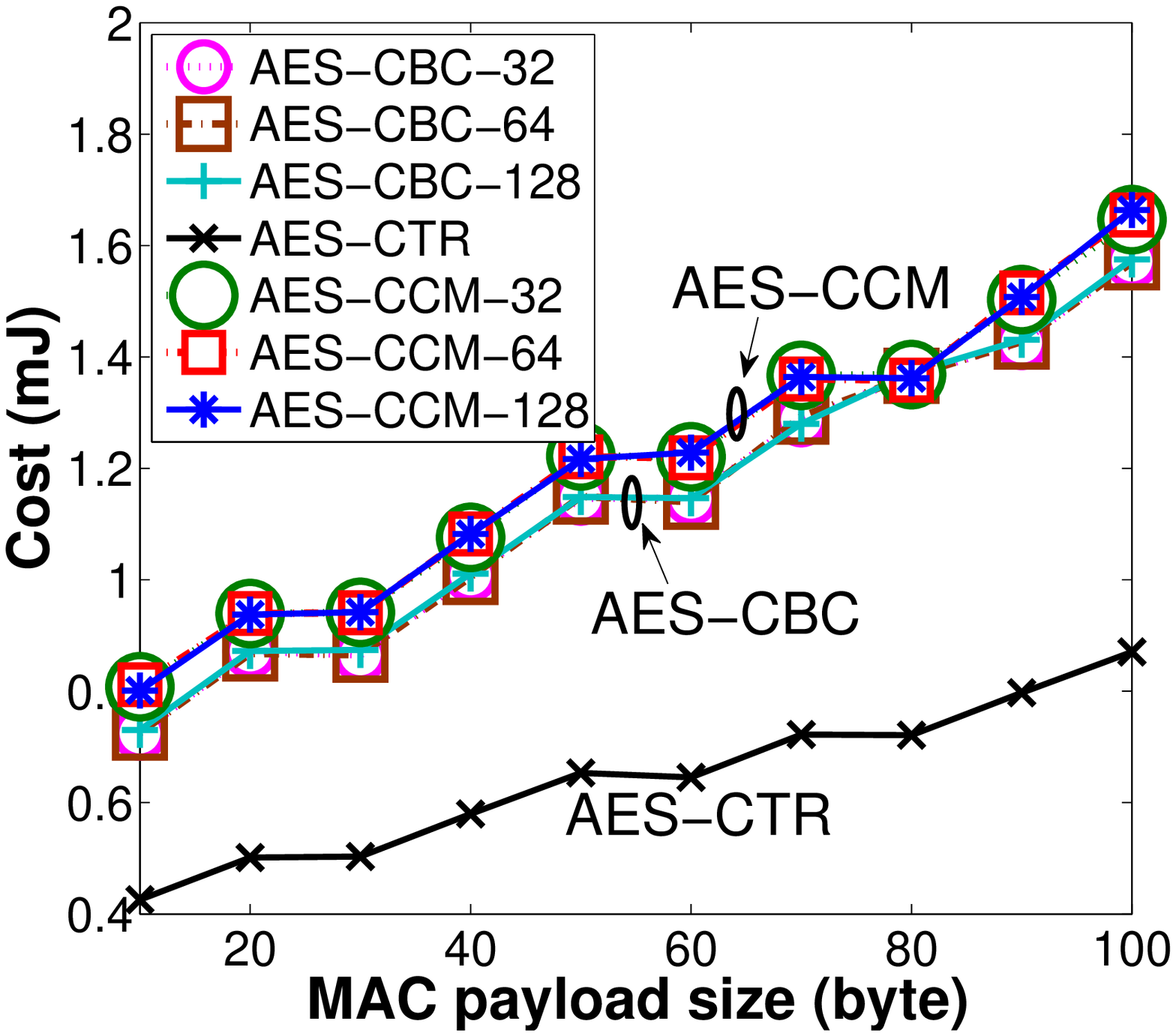}
		\label{fig_pDecCost}
	}
	\hspace{-3mm}
	\subfigure[]{
		\includegraphics[width=1.6in]{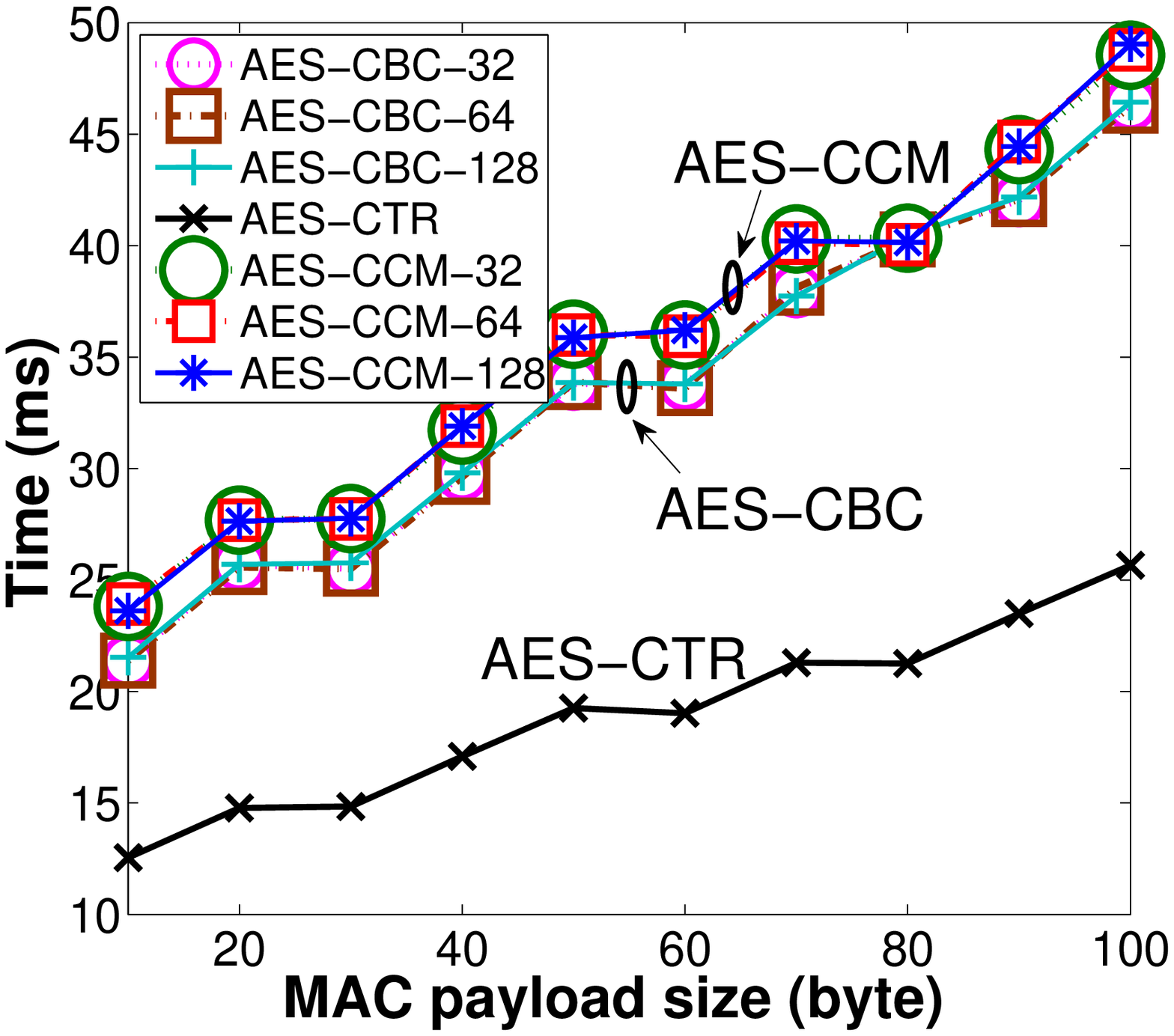}
		\label{fig_pDecTime}
	}
	\hspace{-3mm}
	\subfigure[]{
        \includegraphics[width=1.6in]{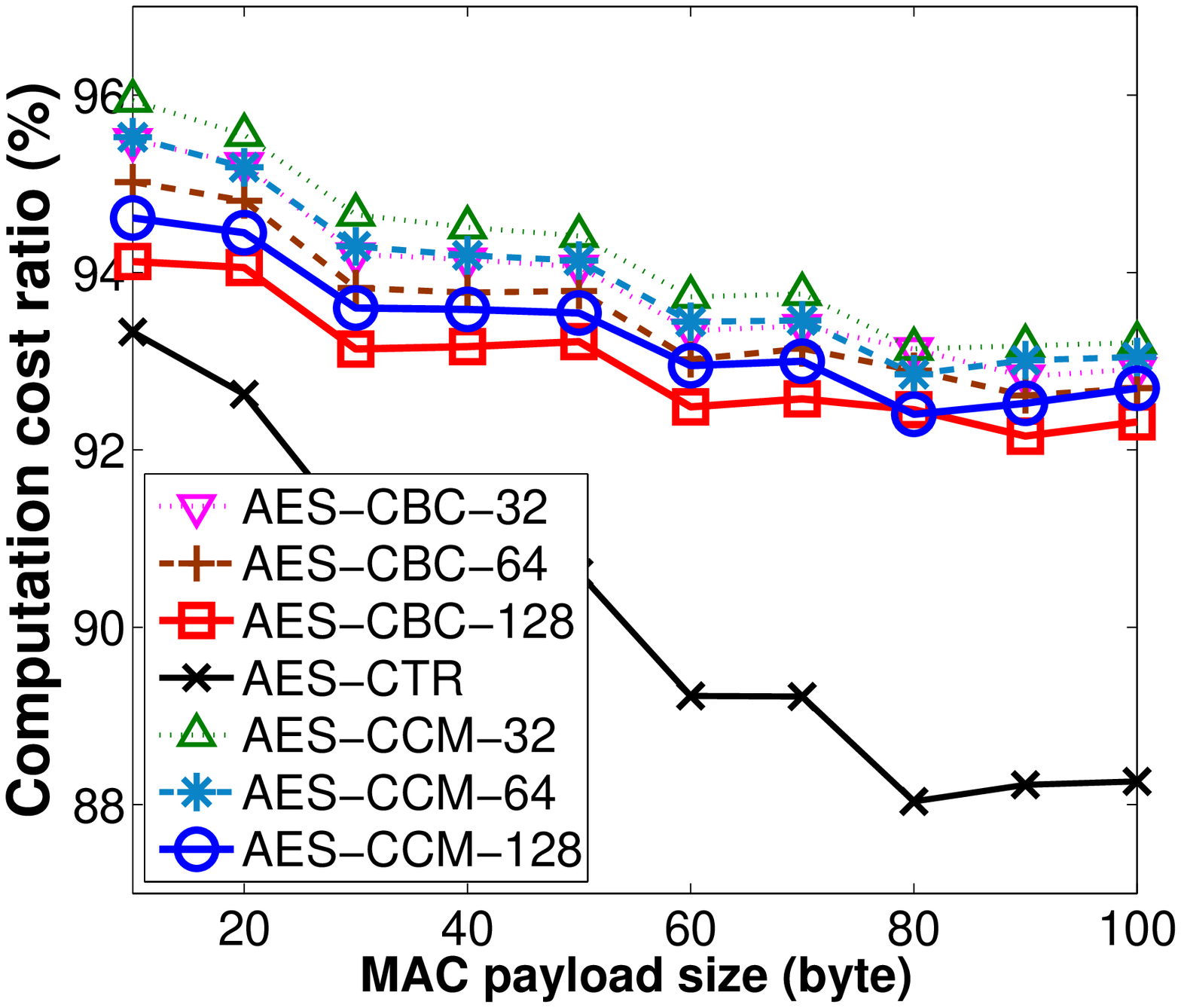}
        \label{fig_pDecCostRatio}
    }
\vspace{-2mm}
	\caption{Energy and time costs. (a) and (b): average energy consumption for encrypting a packet and decrypting a secured packet, respectively. (c): average time (ms) for decrypting and verifying a secured packet. (d): percentage of CPU computational cost in receiving (including decrypting) a secured packet.}	
	\label{Fig_pcost}
\end{figure*}

\subsection{Energy Expenditure for Security}\label{sec:sim:energyperpacket}
We first compare the energy expenditures in encryption and decryption for the security suites listed in Table \ref{tab_level}. To avoid the impact from other factors such as interference and routing, we simulate a two-node scenario in this experiment, i.e., a ghost attacker sends packets to a victim node.  The victim node runs in a duty-cycling mode with fixed duty cycle at 1\%, a typical value for sensor networks \cite{liu2009cmac}. Specifically, the victim node stays in active state for channel listening for 1 ms in every 0.1 second. If there are packets arriving during the active period, the node will process the packets and then switch to sleep if there is still leftover time within the current 0.1-second cycle. If there is no packet received, the node will directly switch to sleep state upon expiration of the active period. Such a mechanism would ensure that the victim device could function properly over one year. To demonstrate the energy-depletion attack, the {\em ghost} attacker sends crafted messages to this victim node every 0.1 second within its active period.

Fig. \ref{Fig_pcost} reports the average computational energy cost and time versus the MAC payload size of the bogus packet for different security suites, where the averages are taken over 25000 bogus packets sent by the attacker. Generally, both encryption and decryption energy costs grow as the payload size increases. Moreover, our simulation results in Fig. \ref{fig_pEncCost} and \ref{fig_pDecCost} demonstrate that the ascending order of the security suites in terms of the average per-packet energy cost (for both encryption and decryption) is as AES-CTR, AES-CBC-MAC, and AES-CCM. This order is reasonable because more data bytes are involved in the XORing operations when the protocol changes from AES-CTR to AES-CCM.

As both the encryption process and the decryption process pad additional 0's to the secured payload in order to partition input data into blocks of 128 bits (16 bytes), the amount of energy expenditure may remain the same for some payload sizes, which is shown in Fig. \ref{fig_pDecCost}. For instance, with AES-CCM-32, a secured packet with MAC payload of 20 bytes will be padded by 10 bytes of 0's so that the resulting string, together with a 2-byte length indicator, has length (in bytes) divisible by 16. However, a MAC payload data of 30 bytes does not need extra 0's. As a result, the two secured packets with different MAC payload sizes consume almost the same computational energy. It is also interesting to observe that, for some MAC payload lengths (e.g., 80 bytes in the figure), the average energy costs for decrypting a secured packet under the AES-CCM and AES-CBC-MAC security suites are quite close. In fact, after padding with necessary 0's, the total numbers of bytes (including secured data and authentication fields) that are input to the XORing operations are very close to each other in these two security suites.

The processing time by the victim node's CPU for decrypting and verifying a secured packet is much longer than that by its transceiver for receiving it. For instance, as shown in Fig. \ref{fig_pDecTime}, with AES-CCM-128, the computation time for decrypting and verifying a packet with 60-byte payload is about 35 ms, while the receiving time is only about 3 ms. As shown in Fig. \ref{fig_pDecCostRatio}, over 88\% of the victim device's energy is spent by its CPU for decrypting crafted packets, which clearly demonstrates the significance of the {\em ghost} attack. We can also observe that AES-CCM-32 introduces the highest CPU cost ratio. Note that the computation cost ratios in all situations slightly decrease when the MAC payload size increases. The reason is that the payload size does not significantly impact the computational energy cost, while it is linearly related to the transceiver's energy consumption in receiving.

\subsection{Node Lifetime under the {\bf \em Ghost} Attack}\label{sec:sim:lifetime}
The {\em ghost} attack can significantly drain the energy of a victim node. For example, we can have a worst-case estimation according to the configuration given in Section \ref{sec:sim:energyperpacket}. In a network where the active period in each 0.1-second cycle is around 1 ms, the ghost attacker sending packet with a payload size of 60 bytes can lead to a busy period around 35 ms.

\begin{figure}[!ht]
\vspace{-2mm}
	\centering
	\includegraphics[width=2.2in]{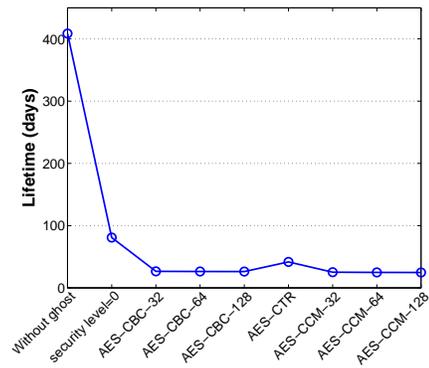}
	\caption{Amount of time (in days) needed to deplete the energy, where the attacker sends bogus packets every 0.1 second.}
	\label{fig_engdep}	
\end{figure}

Fig. \ref{fig_engdep} presents the lifetimes of the victim node in different scenarios including the case without attack and the cases with attacks (operated under the 8 security suites, respectively).  The attack is according to the configuration in Section \ref{sec:sim:energyperpacket} with a MAC payload of 60 bytes. The results in Fig. \ref{fig_engdep} clearly show that the ghost attack significantly shorten the lifetime from over one year to days. Specifically, the lifetime reduces to around 10.2\%, 6.5\% and 6.1\% of the baseline value (i.e., without {\em ghost} attack) under AES-CTR, all the AES-CBC-MAC and all the AES-CCM security suites, respectively. On the other hand, based on the analysis presented in Section \ref{sec:analysis}, we can easily calculate the lifetime radio $\frac{L}{L_0}$ by using the computing time shown in Fig. \ref{fig_pDecTime} and taking into account the sleep cost. The calculated results are 10.9\%, 6.8\% and 6.5\% for AES-CTR, all the AES-CBC-MAC and all the AES-CCM security suites, respectively, which well match the simulation results.

\subsection{Denial of Service Attack}
\begin{figure}[!ht]
%\vspace{-2mm}
\hspace{-4.5mm}
\centering
\begin{minipage}{.22\textwidth}
  \centering
        \includegraphics[width=1.55in, bb=35 175 544 666]{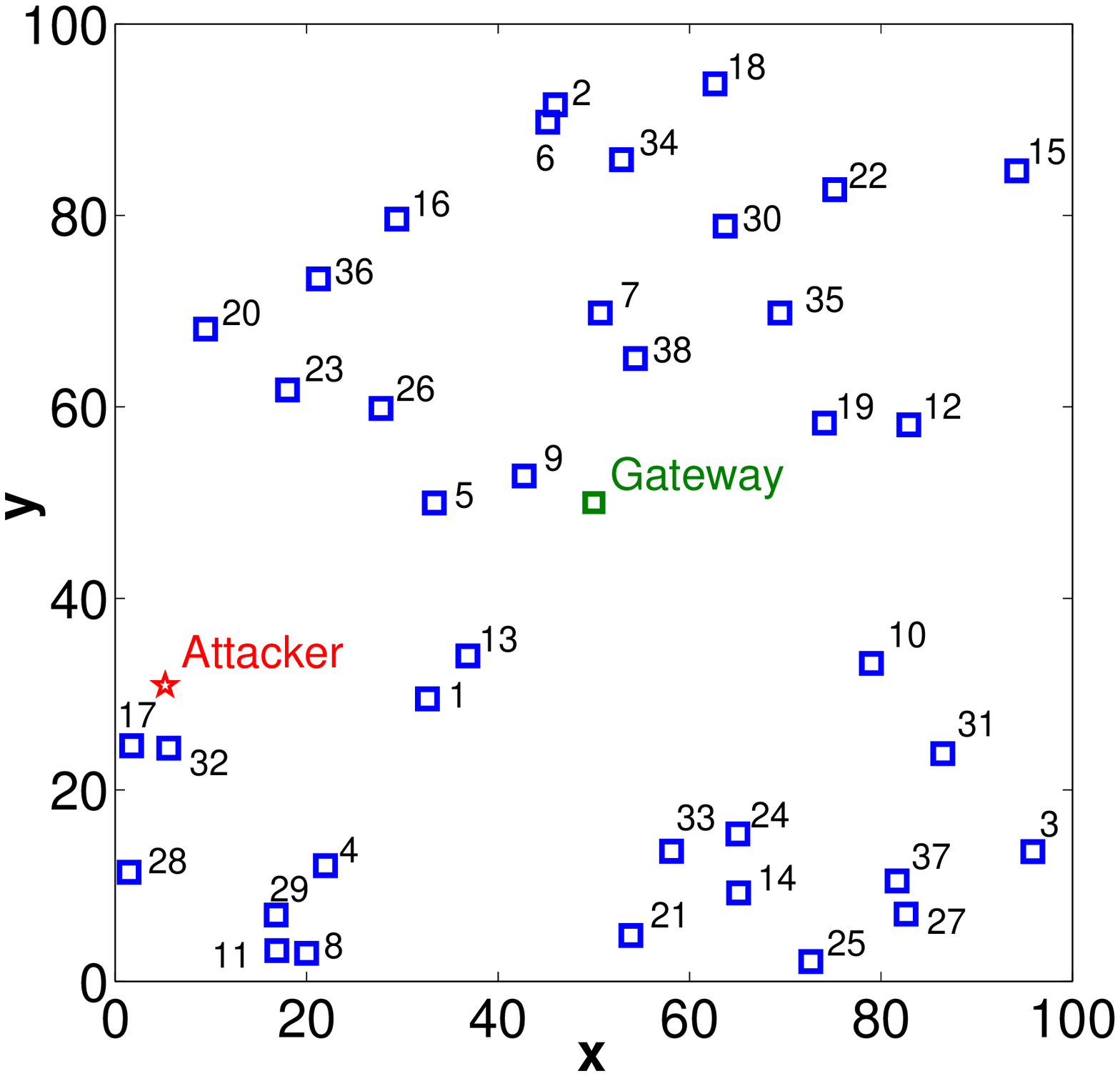}
  \vspace{-4mm}
  \captionof{figure}{Network topology}\label{fig_top}
\end{minipage}%
\hspace{5mm}
\begin{minipage}{.22\textwidth}
  \centering
    	\includegraphics[width=1.55in, bb=36 168 529 658]{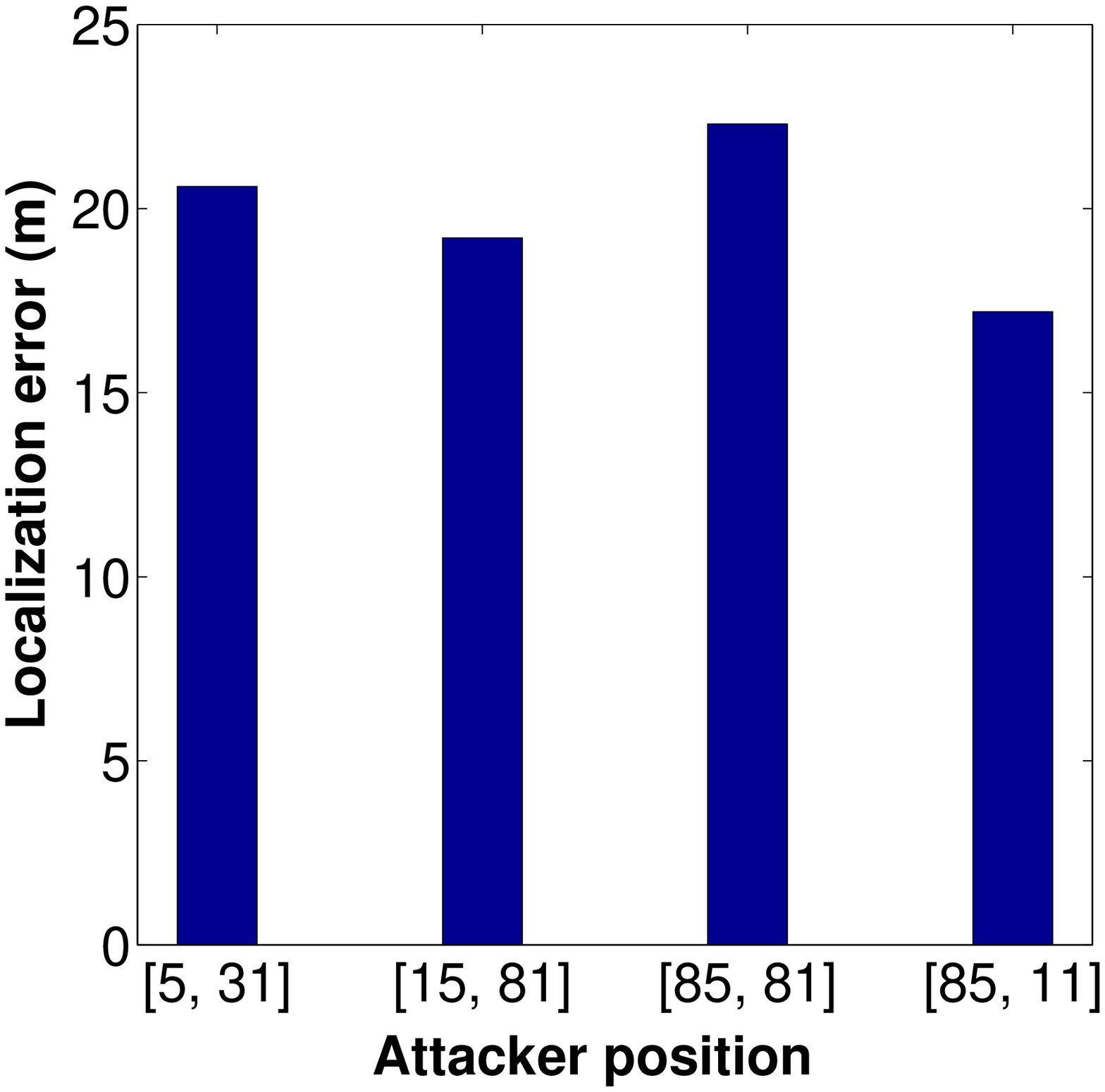}
  \vspace{-4mm}
  \captionof{figure}{Localization error.}\label{fig_locerror}
\end{minipage}
\end{figure}

The following simulations are conducted on a network with $n=38$ legitimate nodes randomly deployed inside a 100$\times$100 $m^2$ area, as shown in Fig. \ref{fig_top}. The gateway node is located at the center of this area. All the nodes (including the attacker node) are assumed to have the same communication range (30$m$) and the same interference range (40$m$), respectively. All the legitimate nodes periodically report data (1 packet/s) to the gateway through multi-hop routes, where we adopt the shortest path routing. The attacker node select node 1 as the victim node and sends bogus packets to it, where the inter-packet sending intervals obey Poisson distribution with 20 ms as the mean. Here we interpret the attack as a DoS attack, according to the discussion in Section \ref{sec:dos}. Simulation results with AES-CTR are reported in the following.

\begin{figure}[!ht]
\vspace{-3mm}
	\centering
	\subfigure[Energy drain speed variation ($\%$).]{
	\centering
    	\includegraphics[width=2.5in, bb=20 213 575 625]{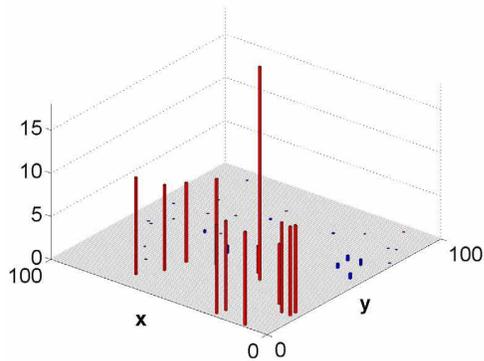}
	    \vspace{-1mm}
        \label{fig_engvar}
    }
	\subfigure[Throughput variation ($\%$).]{
	\centering
    	\includegraphics[width=2.5in, bb=24 213 573 631]{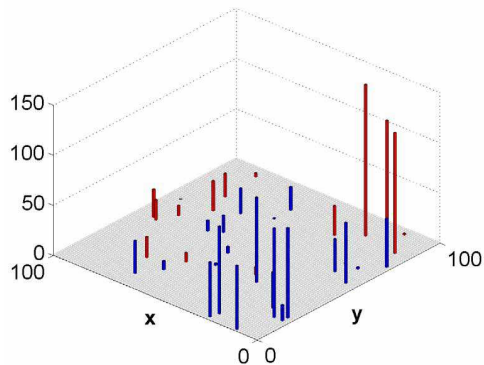}
    	\label{fig_throughput3d}
	}
    \vspace{-2mm}
	\caption{Impact of DoS attack.}	
	\label{Fig_dos}
    \vspace{-2mm}
\end{figure}

Fig. \ref{fig_throughput3d} plots the throughput variation (i.e., the percentage of throughput that is increased/decreased due to the attack) for each node. Note that the red and blue bards mean positive and negative values, respectively. Based on this figure, the DoS attack by the {\em ghost} attacker will change the distribution of the network traffics, and the effects are two-fold. Since the attacker causes node 1 to spend most time on receiving and decrypting crafted packets, the spare capacity that node 1 can provide to others for relaying and also transmitting packets of its own significantly drops. Therefore, nodes, such as node 28, that needs node 1 to relay packets experiences a much higher packet loss rate in the presence of the attack. Although other nodes that receive a crafted packet will directly discard it due to unmatched destination address, there is still bandwidth and energy waste due to channel sensing, packet receiving, and collisions due to the hidden terminal impact \cite{tsertou2008revisiting}. Thus, the throughput of the attacker's neighboring nodes (e.g., node 8, 17 and 32) reduces greatly as shown in Fig. \ref{fig_throughput3d}. Therefore, the attacker can perform denial of service attack at all its neighboring nodes by just targeting at one of them. On the other hand, the presence of the attacker may be beneficial to some other nodes, e.g., node 2, 10, 27, 31 and 37 as shown in Fig. \ref{fig_throughput3d}, which are located far away. For those nodes, because the traffics of some hidden-terminal nodes locating within the attacker's interference range are partly suppressed by the attacker, their successful packet deliver ratios increase in turn.

Fig. \ref{fig_engvar} plots the 3D view of the variation in energy drain speed. We can see that, under the DoS attack, the energy drain speed is accelerated for most of the legitimate nodes. Therefore, the impact of the {\em ghost} attack can be propagated in a multi-hop network such that it can drain the energy of a larger amount of nodes more than that of its neighbors.

\subsection{Effectiveness of Countermeasures}
The performance of the proposed attacker localization method is demonstrated in Fig. \ref{fig_locerror} where the attacker is placed at four different positions. We can see that the localization error is around 20m, which is smaller than the communication range of each legitimate node. For denser networks, a lower localization error may be achieved since the positions of more interfered nodes can be utilized for calculating the attacker's position.

\begin{figure}[htbp]
\vspace{-2mm}
	\centering
	\subfigure[Traffic]{
        \includegraphics[width=1.6in]{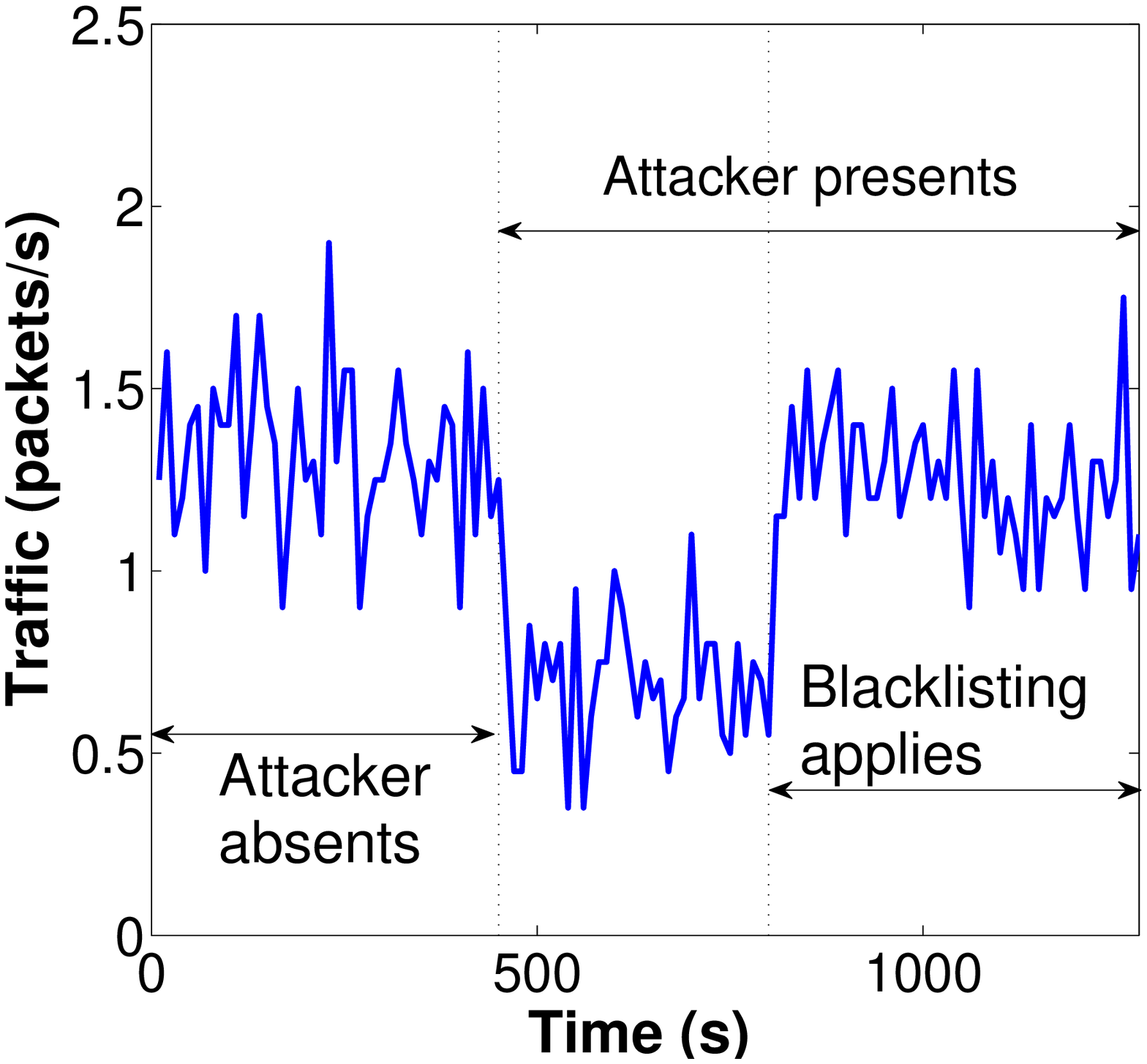}
        \label{fig_bklistingTraffic}
	}
	\hspace{-5mm}
	\subfigure[Energy drain speed]{
        \includegraphics[width=1.6in]{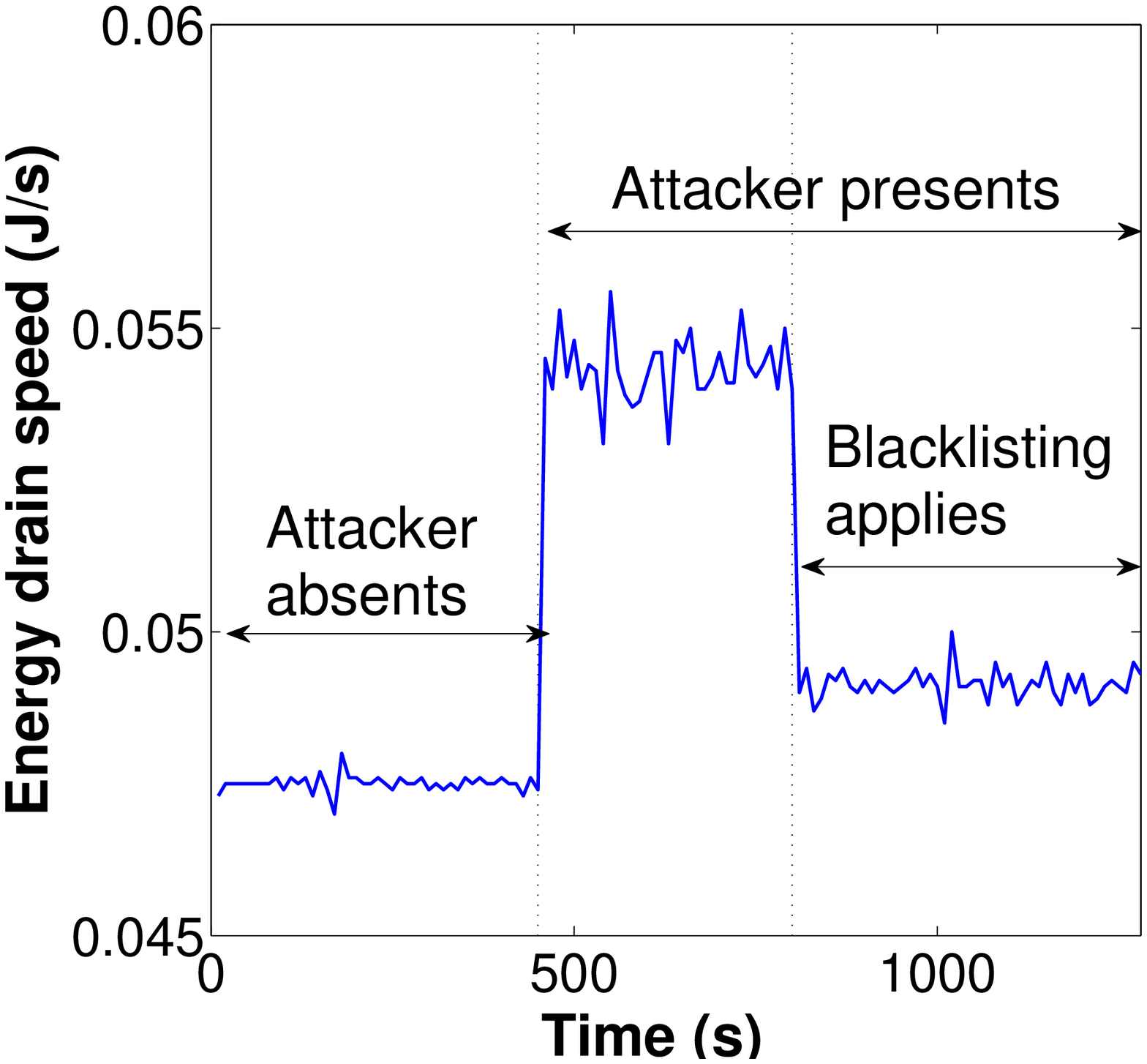}
        \label{fig_bklistingEng}
	}
 \vspace{-1mm}
	\caption{The effect of blacklisting on the victim node's traffic (a) and energy drain speed (b).}	
	\label{Fig_bklisting}
\end{figure}

As shown in Fig. \ref{fig_bklistingTraffic}, the intrusion of the attacker significantly reduces the victim node's traffic, which is then recovered by blacklisting the attacker so that all future crafted packets are directly discarded by the victim node at the MAC layer. In terms of the energy drain speed, similar trends can be observed in Fig. \ref{fig_bklistingEng}. However, the drain speed cannot be fully recovered because the victim node's physical layer energy expenditure for receiving crafted packets is inevitable. In Section \ref{sec:countermeasures}, we also proposed the challenge-response scheme to address both DoS and replay attacks. The traces of the victim nodes' traffic and energy drain rates exhibit similar patterns as shown in Fig. \ref{fig_bklistingTraffic} and \ref{fig_bklistingEng}, and hence are omitted due to the page limit.

\section{Experiments}\label{experiment}
\begin{figure}[!ht]
    \centering
    \includegraphics[width=3.0in, bb=0 314 594 527]{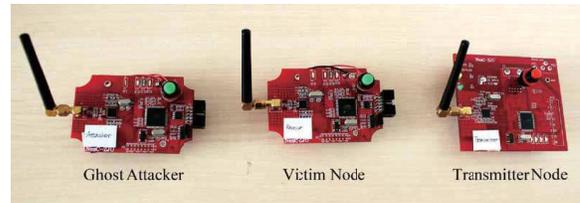}
    \caption{ZigBee nodes in experiments.}
    \label{fig_experiment}
\end{figure}

To further validate the effect of {\em ghost}, we conduct physical experiments based on the ZigBee nodes shown in Fig. \ref{fig_experiment}. Each node has an ATmega128L processor operating with 8MHz frequency, 128K Bytes In-System programmable flash, and a CC2420 RF transceiver compliant with the 2.4GHz IEEE 802.15.4 standard. We develop C programs to activate the embedded security suites of IEEE 802.15.4 and control the security level.

By utilizing an ETCR6000 AC/DC Clamp Leaker with the sampling rate of 2 times per second, the working current under different modes of the victim node is measured and the mean values are shown in Table \ref{workingcurrent}. In the idle state, i.e., MCU is in idle state and the radio is turned off, node's energy is consumed by the basic node operations including the glowing of LED light. When we maintain the MCU executing instructions (e.g., keeping on adding) incessantly, the energy consumption is raised by 7$mA$. Furthermore, when we turn on the radio, the node's working current is remarkably raised. Specifically, if the security suite is implemented into the procedure of transmission, the current reaches around 40$mA$; otherwise it reaches around 32$mA$. The difference of current between with and without security suite, from one aspect, demonstrates the effectiveness of security suite, which happens to be an loophole leveraged by the adversary.

\renewcommand\arraystretch{1.35}
\begin{table}[ht]
\scriptsize
\caption{Node working current.}
\vspace{-4mm}
\center
\begin{tabular}{>{\centering\arraybackslash}m{5.2cm}|c}
\toprule
{\bf Working mode} & {\bf Average current}\\
\hline
Idle & 11$mA$\\
Pure data processing  & 18$mA$\\
Receiving and processing unsecured packets & 32$mA$\\
Receiving and processing secured packets & 40$mA$\\
\bottomrule
\end{tabular}\label{workingcurrent}
\end{table}

\subsection{Single-hop Scenario}
A simple single-hop network consisting of an attack node and a pair of transmitter and receiver/victim nodes is deployed in the experiments. The transmitter and victim nodes, separated by a distance of 0.5m, operate in a low duty-cycling mode. Specifically, they wake up (by turning on their radios) and stay active for roughly 5ms every 50ms. Due to clock drift and the consequent problem of asynchronous communication rendezvous in duty-cycling networks, it is possible that the receiver does not receive anything from the transmitter in one active period. The {\em ghost} attack node is placed close to both the transmitter and victim nodes. It broadcasts 25 bogus packets per second, where each packet has a payload size of 60 Bytes.

We measure the lifetime of the victim node which is powered by a 70-mAh Li-Ion rechargeable battery. We conduct three groups of independent experiments with three batteries. In order to have the same initial energy, the batteries are charged for a same period of time after they have been completely depleted. We assume the {\em ghost} has unlimited resource; thus, it is powered by a D.C. stabilized power supply. The transmitter node's power supply is a normal AA battery which has much higher initial energy than that of the victim node.

\begin{table}[ht]
\scriptsize
\caption{Node lifetime in single-hop scenario. Unit is minute.}
\vspace{-3mm}
\center
\begin{tabular}{c|c|cccccc}
\toprule
 \multicolumn{3}{c}{\multirow{2}{*}{}} & \multicolumn{3}{c}{Battery} & \multicolumn{2}{c}{Average} \\ \cline{4-8}
 \multicolumn{3}{c}{} & \#1 & \#2 & \#3 & {Lifetime} & {Percent} \\ \hline
 \parbox[t]{2mm}{\multirow{4}{*}{\rotatebox[origin=c]{90}{Experimental}}} &
   \multicolumn{2}{c}{\parbox{1.2cm}{No attack}} & 86  & 90  & 100  & 92  & 100\% \\ \cline{2-8}
   & \multirow{3}{*}{\parbox{0.9cm}{Under {\em ghost} attack}} & 011 & 56  & 48   & 65  & 56  & 60.9\% \\ %\cline{2-7}
   & & 100 & 62  & 57  & 71  & 63  & 68.5\% \\ %\cline{2-7}
   & & 111 & 52  & 50  & 60  & 54  & 58.7\% \\ \hline
 \parbox[t]{2mm}{\multirow{4}{*}{\rotatebox[origin=c]{90}{Mapped}}} &
   \multicolumn{2}{c}{\parbox{1.2cm}{No attack}} & 3858  & 4285  & 4458  & 4200  & 100\% \\ \cline{2-8}
  & \multirow{3}{*}{\parbox{0.9cm}{Under {\em ghost} attack}} & 011 & 283  & 215  & 329  & 275  & 6.6\%\\ %\cline{2-7}
  & & 100 & 333  & 275  & 379  & 329  & 7.8\% \\ %\cline{2-7}
  & & 111 & 252  & 227  & 291  & 257  & 6.1\%\\
\bottomrule
\end{tabular}\label{experimentdatacalib}
\end{table}

The experiment results are shown in Table \ref{experimentdatacalib} which clearly justifies the effectiveness of {\em ghost} attack since the node lifetime is obviously shortened under attack. Particularly, with the AES-CCM-128 security suite, the node lifetime is significantly reduced by $39.5\%$, $44.4\%$ and $40.0\%$ with respect to the 3 batteries used, respectively. Moreover, the percentages of lifetime reduction also climbs as the security level increases.

For ease of implementation, the experiments use different settings compared to our simulations in Section \ref{sec:sim:lifetime}, for which reason the results in Table \ref{experimentdatacalib} are different from those shown in Fig. \ref{fig_engdep}. Beside using different battery models and different initial battery power, there are three other major differences between experiments and simulations: 1) A transmitter is introduced to inject packets to the victim node in the experiments. 2) The duty cycle of the victim node is 10\% in the experiments, larger than that in the simulations (which is 1\%). 3) In our simulations, once the victim node enters sleep period, it turns off its radio and switches its CPU to power-save mode. However, in our experiments, the victim node only switches off its radio (while keeping its CPU in idle mode) during its sleep period.

Based on the measured node working current shown in Table \ref{workingcurrent}, the experiment results in Table \ref{experimentdatacalib} are mapped to those shown in Table \ref{experimentdatacalib} by removing the transmitter, using 1\% as the victim node's duty cycle and switching its CPU to power-save mode once it begins sleeping. We can observe that, under the {\em ghost} attack, the node lifetime is remarkably reduced to 6\%$\thicksim$8\% of the lifetime when the attacker is not present. The percentages of lifetime reduction are very close to those observed in the simulations shown in Fig. \ref{fig_engdep}. Therefore, our experiments both confirm the significance of the {\em ghost} attack and validate our methodology of simulations.

\subsection{Multi-hop Scenario}
We construct a multi-hop network whose topology is the same as that displayed in Fig. \ref{fig_multihop}. All the legitimate nodes turn on the radio for 5ms in every 50ms and transmit the packets to the sink node through the fixed path shown in the figure. The ghost attack strategy is implemented in a MicaZ node, which uses the same strategy as in the above single-node scenario to attack Node 2. Each node uses the AES-CCM-128 as its security suite. The experiment results of four representative nodes (Node 1, 2, 3 and 5) are shown in Table \ref{experimentdatamultihop}. We can observe that Node 2, the target of the {\em ghost}, suffers from the fastest speed of energy draining. Its lifetime is shortened as much as $31.6\%$, which basically approaches the performance of the victim node in our single-hop scenario (as compared to the experimental results in Table \ref{experimentdatacalib}). Since Node 1 and 3 also receive the bogus packets, their lifetime is decreased by $15.8\%$ and $16.2\%$, respectively. Since the traffic of Node 1,2 and 3 are suppressed by the {\em ghost}, the throughput of Node 5 increases a little bit, which causes $9.5\%$ reduction of its lifetime. This trend of Node 5 is also demonstrated in Fig. \ref{Fig_multihop} based on our previous analytical model.

\begin{table}[ht]
\scriptsize
\caption{Node lifetime in the multi-hop scenario.}
\vspace{-3mm}
\center
\begin{tabular}{c|cccccc}
\toprule
%\hline
 \multicolumn{2}{c}{\multirow{2}{*}{}} & \multicolumn{3}{c}{Battery Set} & \multicolumn{2}{c}{{Average}} \\ \cline{3-7}
  \multicolumn{2}{c}{} & \#1& \#2& \#3& {Lifetime} & {Percent} \\ \hline
  \multirow{4}{*}{\parbox{0.76cm}{No attack}} & $N_1$ & 85 min & 80 min & 81 min & 82 min & 100\%\\ %\cline{2-7}
   & $N_2$ & 81 min & 78 min & 79 min & 79 min & 100\% \\ %\cline{2-7}
   & $N_3$ & 78 min & 71 min & 72 min & 74 min & 100\% \\ %\cline{2-7}
   & $N_5$ & 88 min & 84 min & 80 min & 84 min & 100\% \\ %\cline{2-7}
  \hline
  \multirow{4}{*}{\parbox{0.76cm}{Under {\em ghost} attack}} & $N_1$ & 74 min & 64 min & 68 min & 69 min & 84.1\%\\ %\cline{2-7}
  & $N_2$ & 53 min & 52 min & 59 min & 54 min & 68.4\% \\ %\cline{2-7}
  & $N_3$ & 60 min & 62 min & 64 min & 62 min & 83.8\%\\ %\cline{2-7}
  & $N_5$ & 77 min & 73 min & 77 min & 76 min & 90.5\% \\ %\cline{2-7}
\bottomrule
\end{tabular}\label{experimentdatamultihop}
\end{table}

\section{Conclusion}\label{conclude}
In this paper, we present a novel and severe attack termed as {\em ghost} in which the malicious one transmits a number of bogus messages to lure the receiving victim device to do the superfluous security-related computations, leading to battery depletion. Our simulation results manifest that by launching this attack, an attacker could easily reduce the lifetime of ZigBee devices from years to days. We also demonstrate that the {\em ghost} attack can further trigger severe DoS and post-depletion attacks. We propose several recommendations on how to withstand the {\em ghost}  and other related attacks in ZigBee networks. Extensive simulations are provided to show the impact of the {\em ghost} attack and the performance of the proposed recommendations. To further validate the effectiveness of {\em ghost} attack, physical experiments were conducted on ZigBee nodes and interestingly, our results show that the lifetime of nodes are significantly impacted with this attack. We believe that the presented work will aid the researchers to improve the security of ZigBee networks further.

\bibliographystyle{IEEEtran}
%\nocite{*}

\bibliography{zigbeeatt}

\end{document}